\documentclass[a4paper,11pt,final]{article}

\usepackage{graphicx,amsmath,amssymb,algorithm,algorithmic,textcomp} 
\usepackage{epsfig}

\begin{document}

\title{\ \\ \LARGE\bf Information content of coevolutionary\\ game landscapes}

\author{Hendrik Richter \\
HTWK Leipzig University of Applied Sciences \\ Faculty of
Electrical Engineering and Information Technology\\
        Postfach 301166, D--04251 Leipzig, Germany. \\ Email: 
hendrik.richter@htwk-leipzig.de. }

\maketitle

\begin{abstract}
Coevolutionary game dynamics is the result of players that may change their strategies and their network of interaction. For such games, and 
based on interpreting strategies as configurations, strategy--to--payoff maps can be defined for every interaction network, which opens up to derive game landscapes. This paper presents an analysis of these game landscapes by their information content. By this analysis, we particularly study the effect of a rescaled payoff matrix generalizing social dilemmas and differences between 
well--mixed and
 structured populations. 

\end{abstract}

\section{Introduction}

Studying evolutionary games can be seen as an attempt to address a long--standing and fundamental problem in Darwinian evolution. How can the two seemingly contradictory observations be reconciled that we have selective pressure entailing
competition between individuals, but at the same time there is wide--spread cooperative and even altruistic behavior among  individuals (and also groups of individuals or even species)?   Evolutionary games set up  mathematical models to discuss the question of whether, when and under what circumstances cooperation may be more advantageous than competition. Such games define individuals having behavioral choices to be players selecting and executing  strategies. By linking the relative costs and benefits of  strategies to payoff (and possibly fitness), we obtain a measure of how profitable a given choice is in evolutionary terms. This discussion crystallizes most prominently into considering so--called social dilemma games, for instance prisoner's dilemma (PD), snowdrift (SD), stag hunt (SH), or harmony (H)~\cite{green12,szabo07}. 

Evolutionary games become coevolutionary if players may not only change their strategies, but also their network of interaction~\cite{perc10}.
Recently, an approach of modeling and analyzing coevolutionary games by dynamic game landscapes has been proposed~\cite{rich16,rich17}, which allows to treat  games within the fitness landscape framework~\cite{rich15,richengel14}. The main advantage of such a game landscape approach is that a strategy--to--payoff map can be defined for every combination of strategies and networks of interaction. Thus, a systematic and quantitative evaluation of the expectable evolutionary dynamics becomes possible. 
In this paper, the approach is extended by studying the information content of game landscapes. Generally speaking, the information content measures the amount of information required to describe such a landscape. For instance, if the landscape is rugged with a larger number of local peaks, the amount of information may be high. If, on the other hand, the landscape is flat or has just a single, smoothly accessible peak, the amount of information can be rather low. Thus, the information content gives a quantitative evaluation of how smooth, neutral or rugged the landscape is. 
These topological features may subsequently be linked to the likelihood of evolutionary paths from a given initial point to target points on the landscape.  
In short, studying the information content is a powerful method of landscape analysis~\cite{mal09,mun15,steer08,vassi00}, which we here apply to coevolutionary games. 

The paper is organized as follows. Sec. \ref{sec:desc} recalls coevolutionary games and particularly discusses  a recently introduced rescaling of the payoff matrix to generalize social dilemmas~\cite{wang15}. Also, modelling interaction networks by $d$--regular graphs is addressed.    Furthermore, it is shown that the strategies of players and coplayers can be interpreted as configurations~\cite{chen13,chen16,rich16,rich17} and given how such configurations relate to strategy--to--payoff maps. In Sec.  \ref{sec:gameland} the  strategy--to--payoff maps are used to define game landscapes for which the information content is analyzed.  Numerical experiments are presented and discussed in Sec.  \ref{sec:exp}. We particularly study how the information content relates to different social dilemmas specified by the rescaled payoff matrix and how the game landscapes vary from a well--mixed population, where every player interacts with all other players, to a structured population, where the interaction matrix imposes restrictions as to who--plays--whom. A summary and concluding remarks end the paper. 

\section{Coevolutionary games, configurations, and strategy--to--payoff maps}  \label{sec:desc}
\begin{figure}[t]
\centering
\includegraphics[trim = 20mm 103mm 40mm 75mm,clip, width=7.0cm, height=4.9cm]{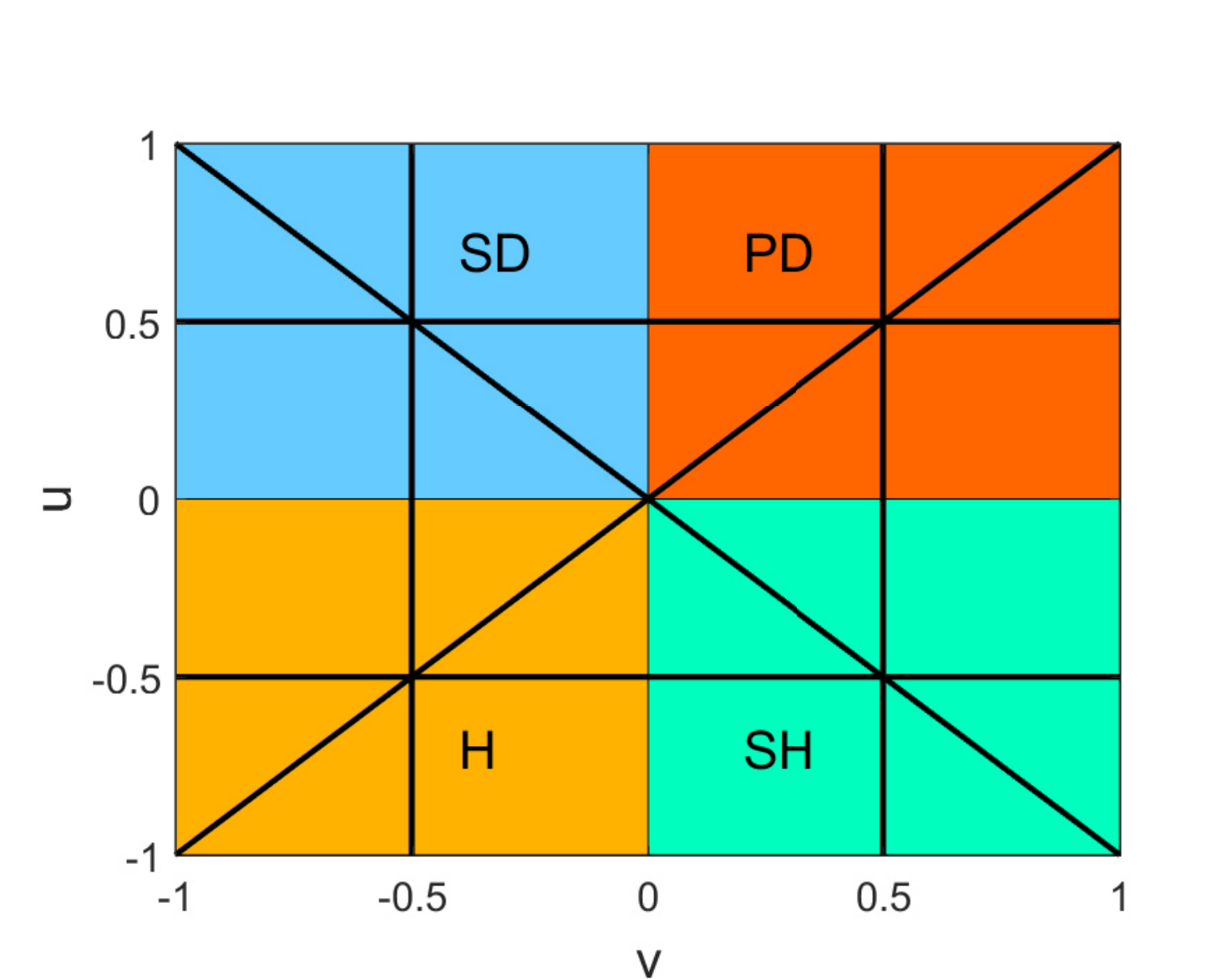} 
\caption{Schematic description of social dilemmas in a $uv$--parameter plane, as defined by the rescaled payoff matrix (\ref{eq:payoff1}). The plane can be divided into four sections (identified by different colors) which correspond to  prisoner's dilemma (PD),  snowdrift (SD), stag hunt (SH),  and harmony (H) games. We define 6 lines bisecting the parameter plane (two diagonal at $u=v$ and $u=-v$, two vertical at $v=-0.5$ and $v=0.5$, and two horizontal at $u=0.5$ and $u=-0.5$), which are traversed in the numerical experiments.}
\label{fig:uv_plane}
\end{figure}

Coevolutionary games of $N$ players are specified by three entities: (i) the payoff matrix, (ii) the network of interaction,  and (iii) the strategy of each player~\cite{perc10,rich16,rich17}.   For a game with two strategies, cooperate ($C_i$) and defect ($D_i$), the pairwise interaction between two players $\mathcal{I}_i$ and $\mathcal{I}_j$, $i\neq j$, (which consequently are mutual coplayers) yields  payoff $(p_i,p_j)$ described by a $2 \times 2$ payoff matrix      
\begin{equation} 
\bordermatrix{~ & C_j & D_j \cr
                  C_i & R & S \cr
                  D_i & T & P \cr} \label{eq:payoff}
\end{equation}
where $T$ is temptation to defect, $R$ is reward for mutual cooperation, $P$ is punishment for mutual defection, and $S$ is sucker payoff for cooperating with a defector. Depending on the values and order of these 4 elements of the payoff matrix (\ref{eq:payoff}), we have different social dilemma games.
Several suggestions have been made to rescale the payoff matrix (\ref{eq:payoff}) by freezing or linearly coupling its elements, which may reduce the four--dimensional parameter space to a two--dimensional plane~\cite{wang15,zuk13}, while preserving frequently--studied social dilemmas such as prisoner's dilemma (PD),   snowdrift (SD), stag--hunt (SH), or harmony (H). Following Wang et al.~\cite{wang15}, we may introduce two scaling parameters $u$ and $v$ to obtain a rescaled payoff matrix
\begin{equation} 
\bordermatrix{~ & C_j & D_j \cr
                  C_i & R & P-(R-P)v \cr
                  D_i & R+(R-P)u & P \cr} \label{eq:payoff1}
\end{equation}
where $u=\frac{T-R}{R-P}$ and $v=\frac{P-S}{R-P}$. We require $R>P$, while $T-R$ and $P-S$ may change sign for having different orders of $(T,R,P,S)$, and thus different social dilemmas. Apparently, matrix (\ref{eq:payoff1}) reduces to matrix (\ref{eq:payoff}) by inserting $u$ and $v$.  However, by varying $u$ and $v$ for $-1 \leq u \leq 1$ and $-1 \leq v \leq 1$, we may traverse a two--dimensional $uv$--parameter plane encompassing all the social dilemmas given above, but also some intermediate forms, see Fig. \ref{fig:uv_plane}. We obtain   SD games for  $0 \leq u \leq 1$  and  $-1 \leq v \leq 0$, PD games for $0 \leq u \leq 1$  and  $0 \leq v \leq 1$, and so on.  Thus, a rescaling by matrix (\ref{eq:payoff1})  significantly eases analyzing the games across social dilemmas. A square in the $uv$--plane generalizes the payoff matrix (\ref{eq:payoff}) and produces a multitude of dilemmas that are significant and interesting in evolutionary game theory. Moreover, Wang et al.~\cite{wang15} have shown that by the rescaling (\ref{eq:payoff1}) fixation properties of the games over the $uv$--plane are fairly robust with respect to the choice of $R$ and $P$.

A second entity to describe coevolutionary games is the interaction network specifying who--plays--whom. According to evolutionary graph theory~\cite{allen14,sha12,szabo07}, each player $\mathcal{I}_i$ belongs to a vertex $i$ of an interaction graph, while an edge connecting vertex $i$ and vertex $j$ shows that the players   $\mathcal{I}_i$ and  $\mathcal{I}_j$ are mutual coplayers.  Algebraically, the interaction network is described by the adjacency matrix $A_I \in [0,1]^{N \times N}$.  The interaction graph having an edge between vertex 
$i$ and vertex $j$ is equivalent to the matrix $A_I$ having an element $a_{ij}=1$, while  $a_{ij}=0$ indicates no interaction between the players   $\mathcal{I}_i$ and  $\mathcal{I}_j$. With no self--play, all $a_{ii}=0$. In a computational approach, interaction networks can be modeled by different classes of   Erd{\"o}s--R{\'e}nyi graphs (E--R graphs)~\cite{hinder15,hinder16,rich16,rich17}. In this paper, we consider interaction networks in which each player has the same number of coplayers. Thus, the interaction network can be described by a special E--R graph, a random $d$--regular graph. Random $d$--regular graphs are particularly suitable for a computational approach involving numerical experiments. There are $\mathcal{L}_d(N)$ different instances of $d$--regular graphs on $N$ vertices that can serve as interaction networks for $N$ players with $d$ coplayers each, thus offering to study the effect of changing the setting as to who--plays--whom. Furthermore, there are fast and efficient algorithms for generating such graphs~\cite{bay10,blitz11}.

 One of the questions we intend to study by the approach presented are the differences in game dynamics between well--mixed populations of players, where every player has all other players as coplayers, and structured populations, where the interaction network allows each player only selected coplayers.  Well--mixed populations have a complete interaction graph. For any $N>1$, there is   $\mathcal{L}_{N-1}(N)=1$  with exactly one complete graph. To address the question of difference, we  relate the results obtained for this graph with results for graphs with $2 \leq d <N-2$ coplayers.

The third entity   needed for describing a coevolutionary game is the strategy vector $\pi(k)$, which comprises of the $N$ strategies $\pi_i(k)$ that players may use in  a given round $k$ of the game. Players may change their strategies. In the following, such a strategy updating is modeled as a Moran process and a frequently used updating rule, birth--death (BD) is studied~\cite{allen14,sha12,szabo07}, without any replacement restrictions. Other updating rules with calculable transition probabilities~\cite{patt15} can be handled within the same framework, for instance death--birth,  imitation, or pair--wise comparison.

The strategy vector $\pi(k)$  can be understood as a configuration of the game~\cite{chen13,chen16,rich16,rich17}. For any finite number of players there is a finite number of configurations. For a game with 2 strategies, there are $\ell=2^N$ configurations, which equals the number of words with length $N$  that can be built over a 2--letter alphabet $\mathcal{A}$, for instance  the strategies cooperate and defect   $\mathcal{A}=\{C_i,D_i\}$, for which we use the binary code  $\mathcal{A}=\{1,0\}$. Consider the example of $N=4$ players with $d=3$ coplayers each. There are $2^4=16$ configurations. For instance, the configuration 
$\pi=(\pi_1\pi_2\pi_3\pi_4)=(0110)$   means that players $\mathcal{I}_2$ and $\mathcal{I}_3$ cooperate, while    $\mathcal{I}_1$ and $\mathcal{I}_4$ defect. From such a configuration, the payoff of every player can be assigned as follows~\cite{chen16}. A specific player, for instance $\mathcal{I}_1$, may choose to cooperate ($\pi_1=1$) or defect ($\pi_1=0$). Its payoff depends on what the coplayers' strategies are. The strategies of the remaining $N-1$ coplayers can also be understood as a configuration, which we may call a {\it coplayer configuration} $\pi_{co}$ to distinguish it from the (full) configuration $\pi$ comprising the strategies of all players. For any player $\mathcal{I}_i$  we may calculate the local frequencies that the coplayers cooperate $\omega_i^1$ or defect $\omega_i^0$. 
Return to example $\pi=(0110)$ and player $\mathcal{I}_1$. The local frequencies of the coplayer configuration $\pi_{co}=(110)$ are $\omega_1^1(110)=2/3$ and $\omega_1^0(110)=1/3$. Note that the local frequencies are independent of the choice of player $\mathcal{I}_1$'s own strategy. Further observe that for the same full configuration 
$\pi=(0110)$, but the perspective of player  $\mathcal{I}_2$, the coplayer configuration is $\pi_{co}=(010)$, which entails different local frequencies ($\omega_2^1=1/3$ and $\omega_2^0=2/3$). 
From the local frequencies of the coplayer configuration $\pi_{co}$ and the payoff matrix (\ref{eq:payoff}), the payoff $p_i$ of a player $\mathcal{I}_i$ can be calculated as \begin{equation} p_i^1(\pi_{co})=R\omega_i^1(\pi_{co}) +S \omega_i^0(\pi_{co}) \label{eq:omega1} \end{equation}   for the player cooperating ($\pi_1=1$) and  \begin{equation} p_i^0(\pi_{co})=T\omega_i^1(\pi_{co}) +P \omega_i^0(\pi_{co}) \label{eq:omega0} \end{equation} for the player defecting ($\pi_1=0$).
Thus, for player $\mathcal{I}_1$ and the coplayer configuration $\pi_{co}=(110)$, we get $p_1^1(110)=(2R+S)/3$ and $p_1^0(110)=(2T+P)/3$. (It may be a matter of convention to normalize the payoff, as here, 
by the number of coplayers $d=3$, but it might be useful  if we are to compare payoffs over varying $d$.) 

 Calculating payoff by Eqs. (\ref{eq:omega1}) and (\ref{eq:omega0}) shows clearly and naturally that the player's reward depends on how frequent a certain strategy is in the population of
coplayers, which is known as frequency--dependence. Also, from  
a computational point of 
view, such a calculation   has some interesting properties. (i) Payoff can be calculated for all $\ell=2^{N-1}$ coplayer configurations $\pi_{co}$ and all players, which gives a complete strategy--to--payoff map, that is, for all full configurations $\pi$. For instance, $p_1^1(110)=p_1(1110)$ and $p_1^0(110)=p_1(0110)$, while  $p_2^1(110)=p_2(1110)$ and $p_2^0(110)=p_2(1010)$, and so on.
(ii) For a complete network of interaction, the calculation is symmetric with respect to players. Any player interacts with coplayers whose strategies are defined by the same coplayer configurations. 
(iii) The calculation is basically bit counting of a binary string, also known as counting the Hamming weight, which has time complexity $\mathcal{O}(N)$. The calculation is further eased by the fact that $\omega_i^1$ and 
$\omega_i^0$ are symmetric by $\omega_i^1+\omega_i^0=d$.  (iv) The calculation of payoff is done in two separable steps, first by computing the local frequencies $\omega_i^1$ and $\omega_i^0$ for every coplayer configuration, second
by  Eqs. (\ref{eq:omega1}) and (\ref{eq:omega0}), which is a linear scaling. In other words, the payoff can be seen as a linear parametrization of the local frequencies. This means, once we have $\omega_i^1$ and $\omega_i^0$, calculating payoff for a parameter plane generalizing
a payoff matrix such as (\ref{eq:payoff1}) becomes numerically  less expensive. Finally, (v) the calculation of payoff is shown for a complete network of interaction (and thus a complete adjacency matrix), but can straightforwardly be extended
to any $d$--regular graph.  

Consider again $N=4$ players but now $d=2$ coplayers. The network of interaction is described by a $2$--regular graph on $4$ vertices, for which there are $\mathcal{L}_2(4)=3$ instances. One of them is $A_I(1)=\left( \begin{smallmatrix} 0 & 0 & 1 & 1 \\ 0 &  0 & 1 & 1 \\ 1 & 1 & 0 & 0\\ 1 & 1 & 0 & 0 \end{smallmatrix} \right)$. With no self--play, the main diagonal of $A_I$ is all $a_{ii}=0$. We may remove these $a_{ii}=0$ and replace the remaining $a_{ij}=0$ by $a_{ij}:=\text{o}$   to obtain a coplayer adjacency matrix $A_{I_{co}} (1)=\left( \begin{smallmatrix} \text{o} & 1 & 1 \\ \text{o} &  1 & 1 \\ 1 & 1 & \text{o} \\ 1 & 1 & \text{o}  \end{smallmatrix} \right)$.  
The same yields
 $A_{I_{co}}(2)=\left( \begin{smallmatrix} 1 & \text{o} & 1 \\ 1 &  1 & \text{o} \\ \text{o} & 1 &  1\\ 1 & \text{o} & 1  \end{smallmatrix} \right)$, $A_{I_{co}}(3)=\left( \begin{smallmatrix} 1 & 1 & \text{o} \\ 1 &  \text{o} & 1 \\ 1 & \text{o} &  1\\ \text{o} & 1 & 1  \end{smallmatrix} \right)$ for the remaining two instances. The $i$--th row of the coplayer adjacency matrix describes if there is interaction of player $\mathcal{I}_i$ with the remaining players ($a_{ij}=1$) or not  ($a_{ij}=\text{o}$). Thus, by observing  $1\text{o}=0\text{o}=\text{o}$, we get for player $\mathcal{I}_1$, $A_{I_{co}} (1)$  and $\pi_{co}=(110)$ the element--wise product $(110) \circ (\text{o}11)=(\text{o}10)$. The Hamming weight  discarding the element $(\text{o})$ in the string $(\text{o}10)$  gives the local frequency to cooperate as $\omega_1^1=1/2$ and the local frequency to defect as $\omega_1^0=1/2$.   Note that for the same player $\mathcal{I}_1$ and the same coplayer configuration, the local frequencies might be different for another interaction network. For $A_{I_{co}}(3)$, we obtain   $(110) \circ (11\text{o})=(11\text{o})$ with $\omega_1^1=1$ and  
 $\omega_1^0=0$. Thus, for coevolutionary games with varying interaction  networks, the payoff of a player $\mathcal{I}_i$  may depend not only on its own strategy, the coplayer configuration $\pi_{co}$ summarizing the strategies of the coplayers, but also on  the adjacency matrix $A_I$. According to Eqs. (\ref{eq:omega1}) and (\ref{eq:omega0}) the payoffs are 
$p_1^1(110)=(R+S)/2$ and $p_1^0(110)=(T+P)/2$ for   $A_{I}(1)$ and $A_{I}(2)$, but $p_1^1(110)=R$ and $p_1^0(110)=T$ for   $A_{I}(3)$. 
The local frequencies and thus the payoff may (but does not have to) vary over different networks of interaction. Each network of interaction may have its own strategy--to--payoff map.  Also for varying interaction networks, calculating payoff can be separated into two steps: (i) computing the local frequencies for all coplayer configurations, all players and the $A_I$'s, and (ii)  using Eqs. (\ref{eq:omega1}) and (\ref{eq:omega0})  and any payoff matrix for a linear scaling.

\section{Game landscapes and their information content} \label{sec:gameland}
The previous section has shown how for all players and each network of interaction a strategy--to--payoff map can be constructed. We now define game landscapes from these strategy--to--payoff maps~\cite{rich16,rich17}. Therefore, we calculate the fitness $f_i(\pi)$  from the payoff $p_i(\pi)$ for all strategy configurations $\pi$ and each player $\mathcal{I}_i$ by $f_i(\pi)=1+\delta p_i(\pi)$ with the intensity of selection $\delta>0$. We further observe that according to the Moran process only one player may change its strategy at a given point of time. Thus,  for $N$ players with   configuration $\pi=(\pi_1\pi_1\ \ldots \pi_N)$, there are $N$ possibilities for strategy updating. Put another way, each configuration has $N$ neighbors, and for a binary representation of configurations, the neighborhood structure  is Hamming distance of $1$, denoted by $\mathcal{H}_d^1$. Finally, we may combine all configurations $\pi$ to form a strategy configuration space $\Pi$. Hence,
for each interaction network specified by an adjacency matrix $A_I$ and each player, we get a landscape $\Lambda_\Pi^i=(\Pi,\mathcal{H}_d^1,f_i(\pi))$  that maps configurations $\pi \in \Pi$ with a neighborhood defined by Hamming distance $\mathcal{H}_d^1$ to fitness $f_i:\Pi \rightarrow \mathbb{R}$. We may further process these landscapes $\Lambda_\Pi^i$ to incorporate different strategy updating schemes, for instance BD. Based on transitions probabilities of the updating scheme~\cite{patt15}, a fitness function of a game landscape with BD updating can be obtained by
$f(\pi):= \frac{1} {\frac{1}{2} \left(1+\exp{\left(  \frac{ \alpha}{N} \sum_{i=1}^{N}   \phi_i(\pi)  \right) }  \right) }$
where $\phi_i(\pi)=f_i(\pi)/ \sum_{\ell=1}^{2^N}  f_i(\pi(\ell))$ and $\alpha>0$ is a sensitivity weight, see~\cite{rich17} for details.

Suppose we intend to analyze this game landscape in terms of neutrality, smoothness and ruggedness by making a finite number of observations that each records the fitness of a configuration. It has been shown that these observations are particularly useful if they are the result of a random walk as such a walk may capture fitness relations between neighboring configurations across the landscape~\cite{mun15,vassi00}. Thus, the landscape analysis is based on a sequence  \begin{equation}  \mathcal{S}=(f(\pi_0),f(\pi_1),\ldots,f(\pi_{T-1}) ) \label{eq:seq} \end{equation} of walk length $T$ where each $x_i$ and $x_{i+1}$, $i=0,1,\ldots, T-2$ are neighboring configurations according to $\mathcal{H}_d^1$. For calculating the information content we first convert of sequence $\mathcal{S}$  into a symbol sequence  \begin{equation}  S_{ic}=s_0s_1\ldots,s_{T-2}\label{eq:seq1} \end{equation} of length $T-1$, where  the symbols $s_i$ are 
taken from the set $\mathbb{S}= \{-1,0,1\}$. Applying the fitness sequence  (\ref{eq:seq}), the symbols $s_{i} \in \mathbb{S}$ are
calculated   by
\begin{equation}s_i(\epsilon)=\left\{ \begin{array}{rcccc} -1,
& \quad \text{if} \quad& f(\pi_{i+1})-f(\pi_{i}) & < & \epsilon \\ 0, & \quad \text{if} \quad &  |f(\pi_{i+1})-f(\pi_{i})|&\leq & \epsilon\\
1, & \quad \text{if}  \quad& f(\pi_{i+1})-f(\pi_{i})&> & \epsilon
\end{array} \right.
\end{equation}
for a fixed $\epsilon \in [0,E]$, where $E$ is the maximum difference
between two fitness values. 
The parameter $\epsilon$ defines the sensitivity of the symbol sequence
$S_{ic}(\epsilon)$  accounting for differences in fitness.  For example, 
if $\epsilon=0$, the sequence $S_{ic}(\epsilon)$ contains the symbol $s_i=0$ only for a
random walk on a strictly flat area. Hence, $\epsilon=0$
discriminates very sensitively between increasing and decreasing
fitness values. By contrast, for $\epsilon=E$, the string only
contains the symbol $s_i=0$, which makes any evaluation nonsensical. Thus, a fixed value of $\epsilon$ with
$0<\epsilon<E$ defines a sensitivity level with respect to the information gained from the
landscape structure.
For defining the information content of the landscape, the
distribution of subblocks of length 2, $s_{i}
s_{i+1}$, $i=0,1,\ldots T-3$, within the sequence
(\ref{eq:seq1}) is analyzed. These subblocks stand for local patterns in
the landscape. The probability of the occurrence of the
pattern $ab$ with $a,b \in
\mathbb{S}$ and $a \neq b$ is denoted by $p_{ab}$. As the set $\mathbb{S}$ consists of 3
elements, we find $6$ different subblock $s_{i}
s_{i+1}=ab$ with $a \neq b$
within the sequence $S_{ic}(\epsilon)$. From their probabilities
and a given sensitivity level $\epsilon$  the entropic
measure
\begin{equation} h_{ic}(\epsilon)=- {\underset{a, b \in
\mathbb{S} \atop a \neq b}{\sum}} p_{ab}  \log_6  p_{ab}
\label{eq:infcont}
\end{equation}
is calculated, which is called  information content of the fitness landscape~\cite{mun15,vassi00}.
Note that by taking the logarithm with
the base $6$ in Eq. (\ref{eq:infcont}), the information content is scaled to the interval
$[0,1]$.
It has been argued that the information content (\ref{eq:infcont}) may be a good measures of ruggedness, but may not well capture smoothness and flatness of a landscape~\cite{vassi00}. As an additional measure, the partial information content $m_{ic}$ has been suggested~\cite{mun15,vassi00}. For calculating this measure, we remove from the symbol sequence (\ref{eq:seq1}) any symbol $s_i=0$ and any repeating symbols $s_i=s_{i+1}$ to obtain a ``cleared up'' sequence $S_{ic}^*$. A length comparison between  $S_{ic}^*$  and the initial symbol sequence $S_{ic}$ gives the partial information content $m_{ic}$: 
\begin{equation} m_{ic}(\epsilon)=\frac{| S_{ic}^* |}{| S_{ic} |}. \label{eq:infcontpar} \end{equation}
 In numerical experiments, it can be observed that the values of $h_{ic}$ and $m_{ic}$ may vary over $\epsilon$, see also the numerical experiments given in Sec.  \ref{sec:exp}.  It has been shown~\cite{mal09,mun15,steer08} that information about landscape ruggedness and smoothness are most meaningful  for the maximum information content $h_{ic}={\underset{\epsilon} {\max}} \:  h_{ic}(\epsilon)$ and the maximum partial information content $m_{ic}= \left. m_{ic}(\epsilon)\right|_{\epsilon=0}$.

\begin{figure*}[t]

\includegraphics[trim = 23mm 100mm 40mm 90mm,clip, width=4.3cm, height=4cm]{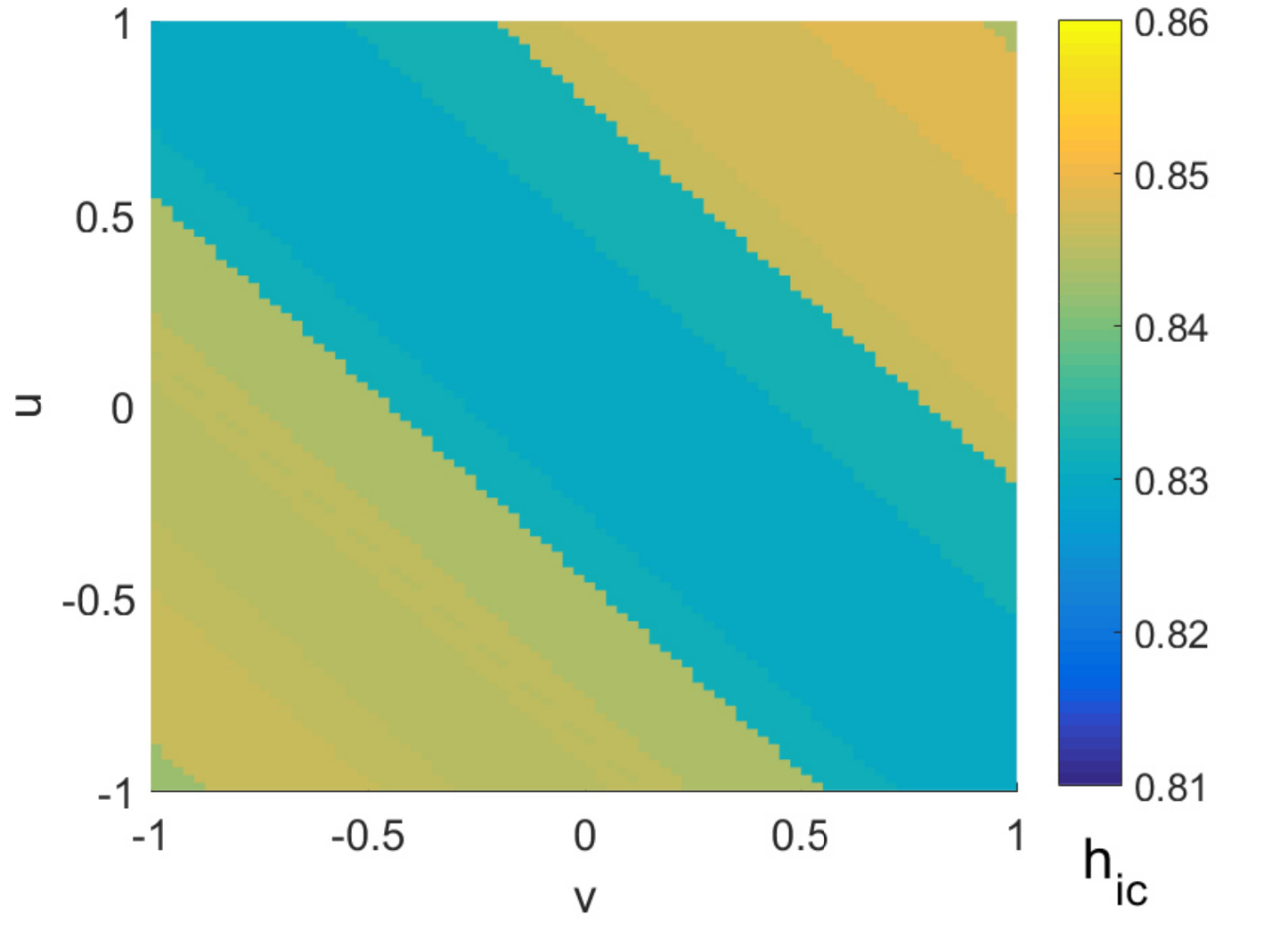} 
\includegraphics[trim = 23mm 100mm 40mm 90mm,clip, width=4.3cm, height=4cm]{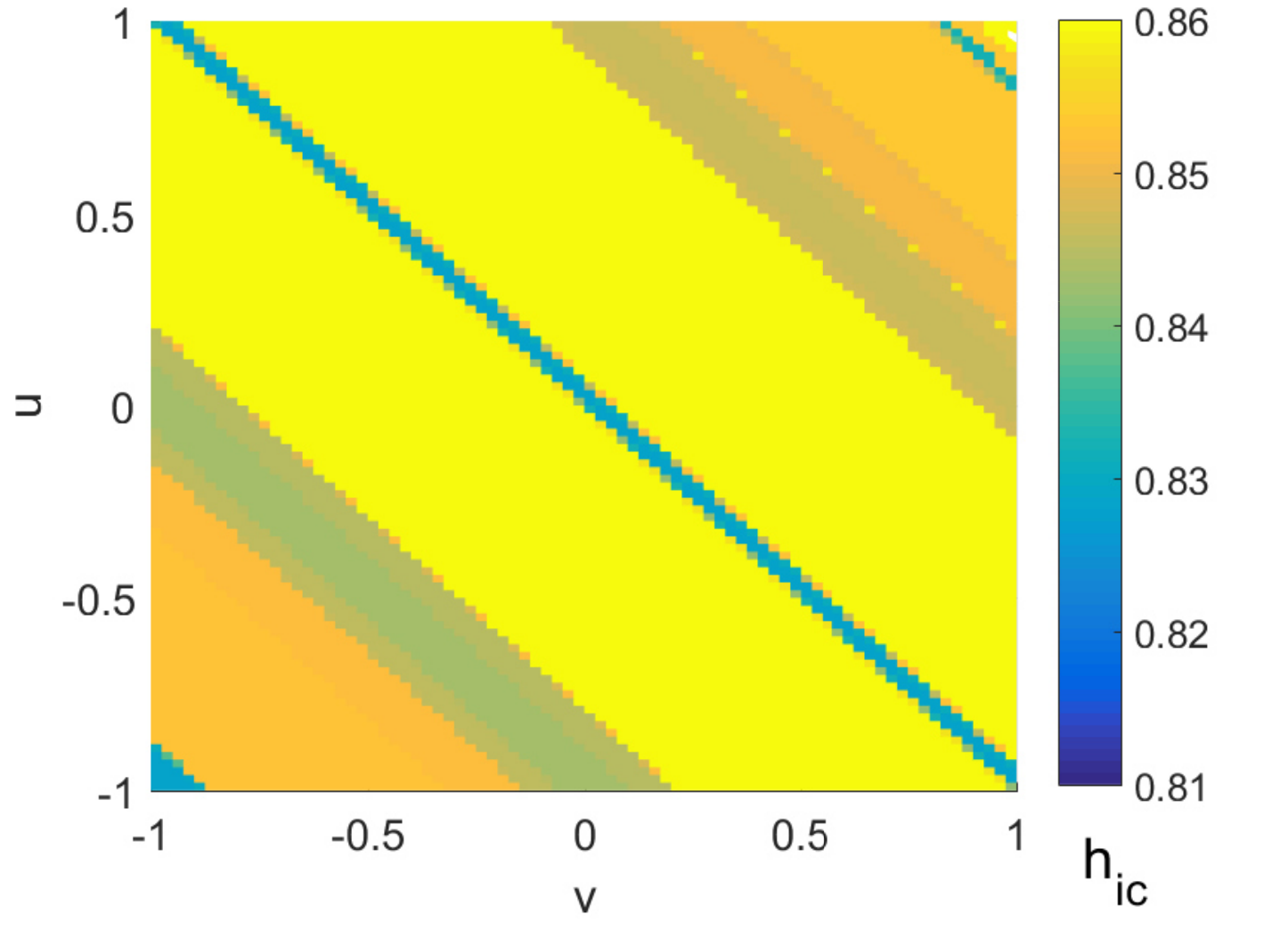} 
\includegraphics[trim = 23mm 100mm 40mm 90mm,clip, width=4.3cm, height=4cm]{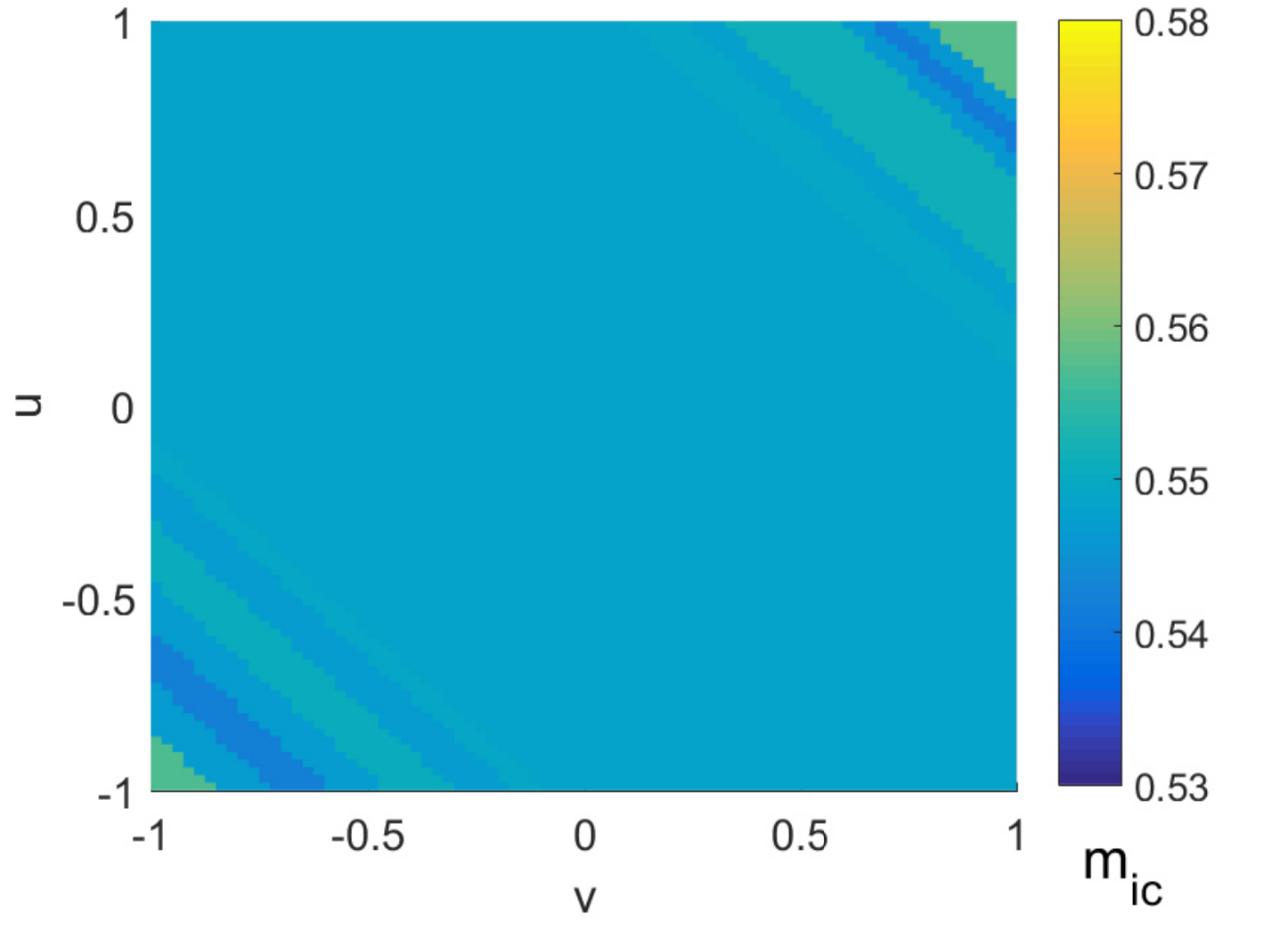} 
\includegraphics[trim = 23mm 100mm 40mm 90mm,clip, width=4.3cm, height=4cm]{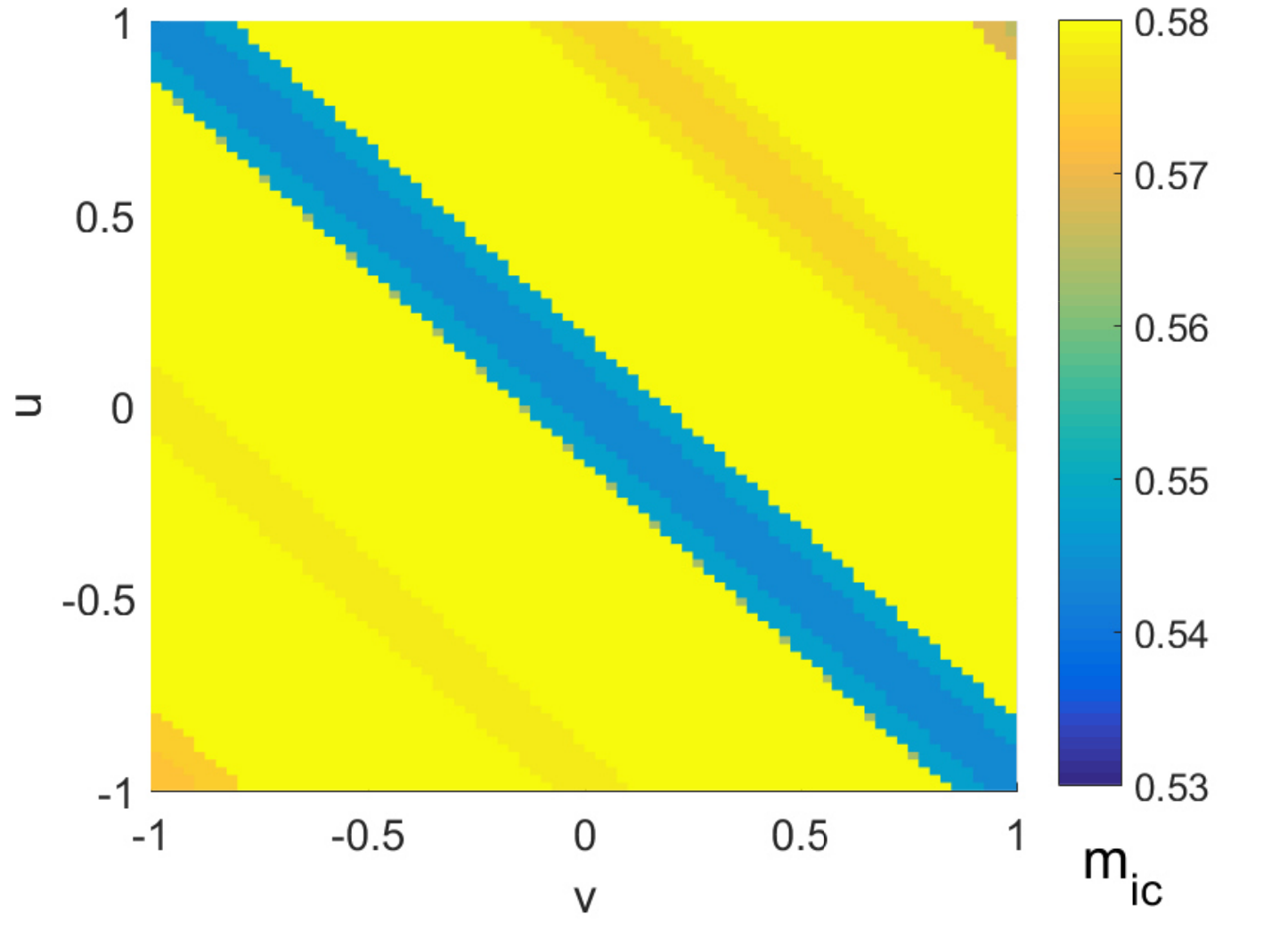}
 
\hspace{1cm} (a) \hspace{4cm} (b) \hspace{4cm} (c) \hspace{4cm} (d) 
\caption{Information characteristics over the $uv$--parameter plane generalizing payoff matrices, see Fig. \ref{fig:uv_plane}, and encompassing relevant social dilemmas. (a)--(b): maximal information content $h_{ic}$,  see Eq. (\ref{eq:infcont}); (c)--(d):  maximal partial information content  $m_{ic}$, see Eq. (\ref{eq:infcontpar}).  The underlying landscapes are based on:  (a), (c): a well--mixed population with $N=6$ and $d=5$ (and thus $\mathcal{L}_5(6)=1$); (b), (d): a structured population with $N=6$ and $d=2$ (and $\mathcal{L}_2(6)=70$) for which a typical network is shown.}
\label{fig:n0}
\end{figure*}

The procedure for calculating $h_{ic}$ and $m_{ic}$ may also be applied to coevolutionary games. Consider we have a series of dynamic instances of $\Lambda_\Pi$ imposed by varying interaction networks described by adjacency matrices $A_I(\kappa)$. Interpreting these dynamic instances as dynamic landscape~\cite{rich17}, a series of landscape measures can be obtained. Thus, we get $ h_{ic}(\kappa)$ and  $m_{ic}(\kappa)$.  Next section, numerical experiments are given and discussed that analyze the information content of coevolutionary game landscapes.

\section{Numerical results and discussion} \label{sec:exp}

The numerical experiments study coevolutionary game landscapes specified by the rescaled payoff matrix (\ref{eq:payoff1}) and varying interaction networks. Therefore, a set of adjacency matrices $A_I(\kappa)$ with given order and degree are generated algorithmically~\cite{bay10,blitz11}. The number of different interaction networks may be very large, even for a moderate number of players. For instance, for $N=12$ players with  $d=2$ coplayers each, we have $\mathcal{L}_2(12)=34.944.085$ different networks~\cite{rich17}. Thus, it is not feasible to numerically evaluate all networks, which is dealt with by the magnitude of the set taken into account be bounded from above by $G=8.500$. For $\mathcal{L}_d(N)<G$, the complete set is used. The game landscapes are calculated for the sensitivity weight $\alpha=5$ and the intensity of selection $\delta=0.25$. The landscape analysis is done by a random walk of length $T=10.000$. The results are averaged over $50$ independent walks. Preliminary experiments have shown no dependence on the initial state of the walk, which makes it reasonable to assume that the landscapes are isotropic.  We vary the $uv$--parameter plane of the rescaled payoff matrix (\ref{eq:payoff1}) by $-1 \leq u \leq 1$ and $-1 \leq v \leq 1$ to encompass all relevant social dilemmas, see Fig. \ref{fig:uv_plane}, and use the parameter  $(R,P)=(1,0)$.

Fig. \ref{fig:n0} shows maximal  information content $h_{ic}$ and maximal partial information content $m_{ic}$ for a well--mixed population ($N=6$, $d=5$) and a typical structured population ($N=6$, $d=2$) over the $uv$--parameter plane. We find that the values for $h_{ic}$ and $m_{ic}$ vary more for $d=2$ (structured) than for $d=5$ (well--mixed). This is a typical result and can also be found for all other interaction networks with $N=6$ and $d=2$ coplayers. For both $d=2$ and $d=5$, the values group along diagonals from north--west to south--east in the parameter plane.  This is in line with previous results showing that for instance fixation properties such as the equilibrium fraction of cooperators also have such a diagonal grouping~\cite{wang15}. In addition, it is conspicuous that the values have much similarity between the south--west and the north--east, but also between the north--west and the south--east corner. This seems paradoxical at the first glance, as these corners represent substantially different social dilemmas, see also Fig. \ref{fig:uv_plane} which shows H and PD games along the south--west--to--north--east diagonal, and   SD and SH games along the  north--west--to--south--east diagonal. However, a fitness landscape is basically a tool for analyzing evolutionary dynamics. Thus, the results about game landscapes in Fig. \ref{fig:n0} relate to game dynamics. Regarding game dynamics, there is similarity between H and PD on the one hand, and SD and SH on the other, and to a much lesser degree between for instance H and SH, or PD and SD. The game dynamics of H and PD games is characterized by monomorphic Nash equilibria where players either all  cooperate (H) or all defect (PD). In other words, in H games cooperation simply dominates defection, while in PD games defection dominates cooperation. By contrast,  SD and SH games possess a more complex composition of Nash equilibria.  For SH, there are bi--stable equilibria  with  players either all cooperating or all defecting.  For SD, there are three polymorphic equilibria with either players choosing opposite strategies ($\pi_i=1$ and $\pi_j=0$ or $\pi_i=0$ and $\pi_j=1$) or players randomly switching between cooperating and defecting. Put another way, in SH and SD games cooperation and defection may alternate or coexist.

\begin{figure}[t]

\includegraphics[trim = 23mm 102mm 40mm 100mm,clip, width=4.3cm, height=4cm]{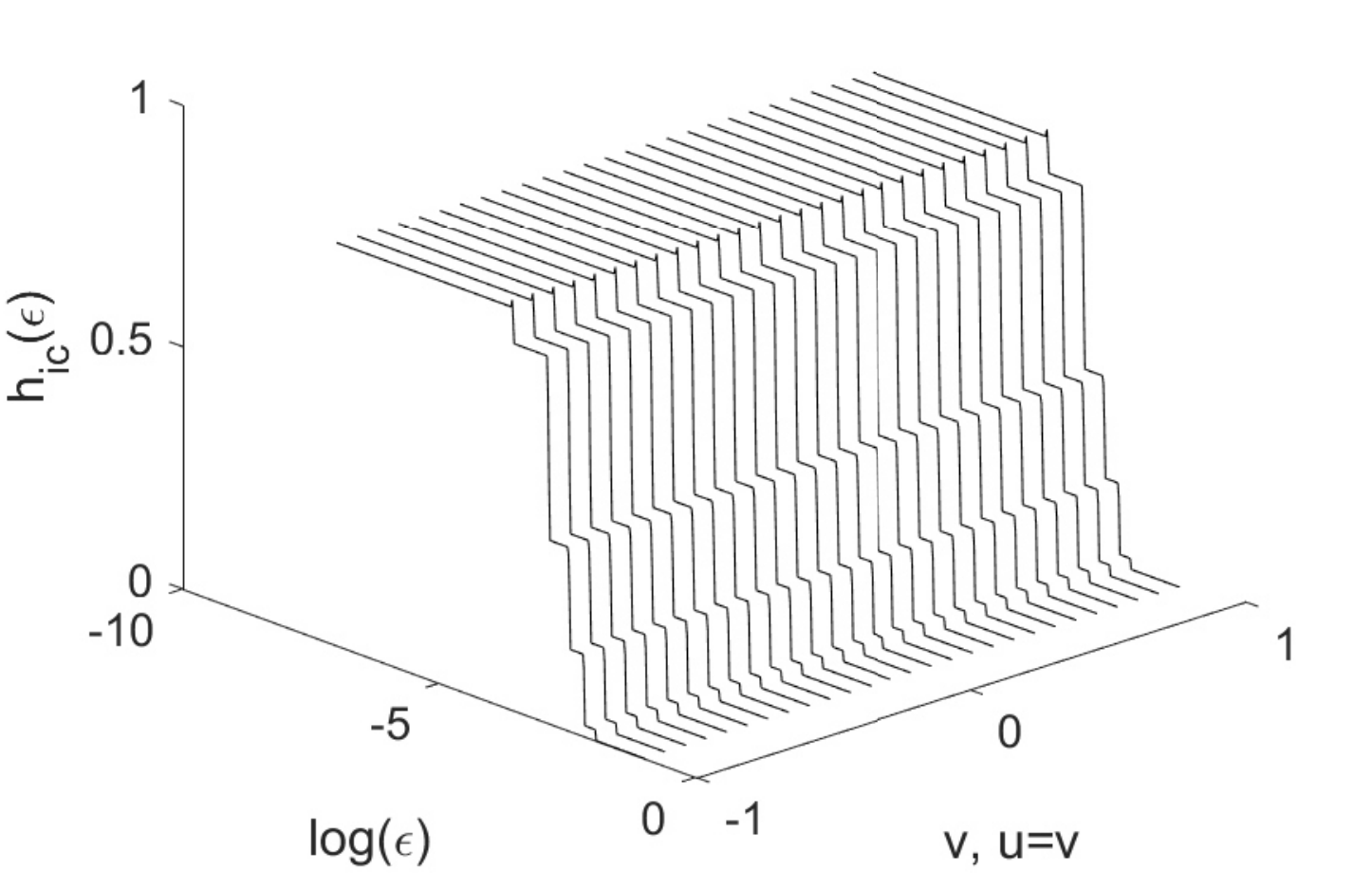} 
\includegraphics[trim = 23mm 102mm 40mm 100mm,clip, width=4.3cm, height=4cm]{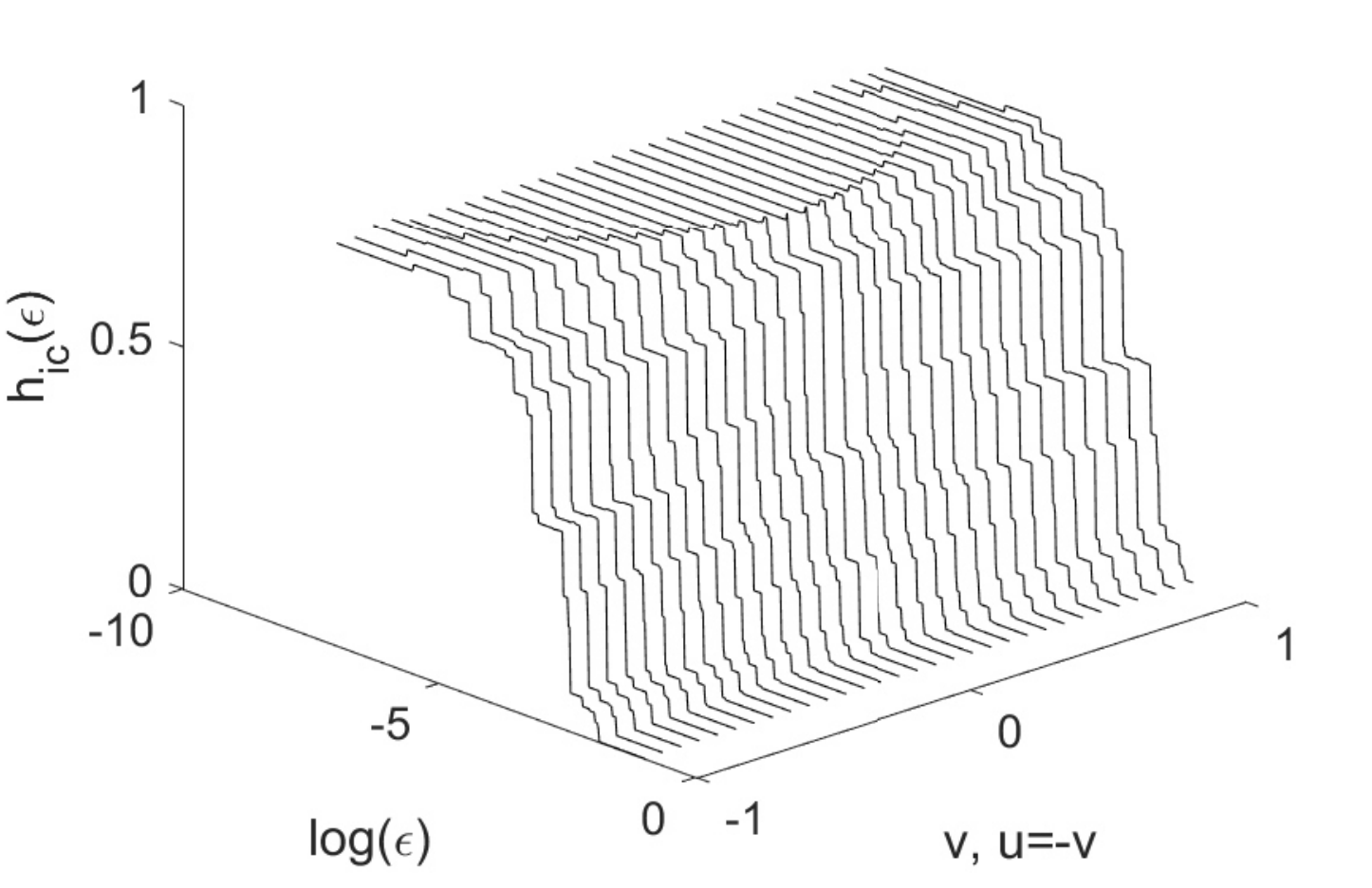} 
\includegraphics[trim = 23mm 102mm 40mm 100mm,clip, width=4.3cm, height=4cm]{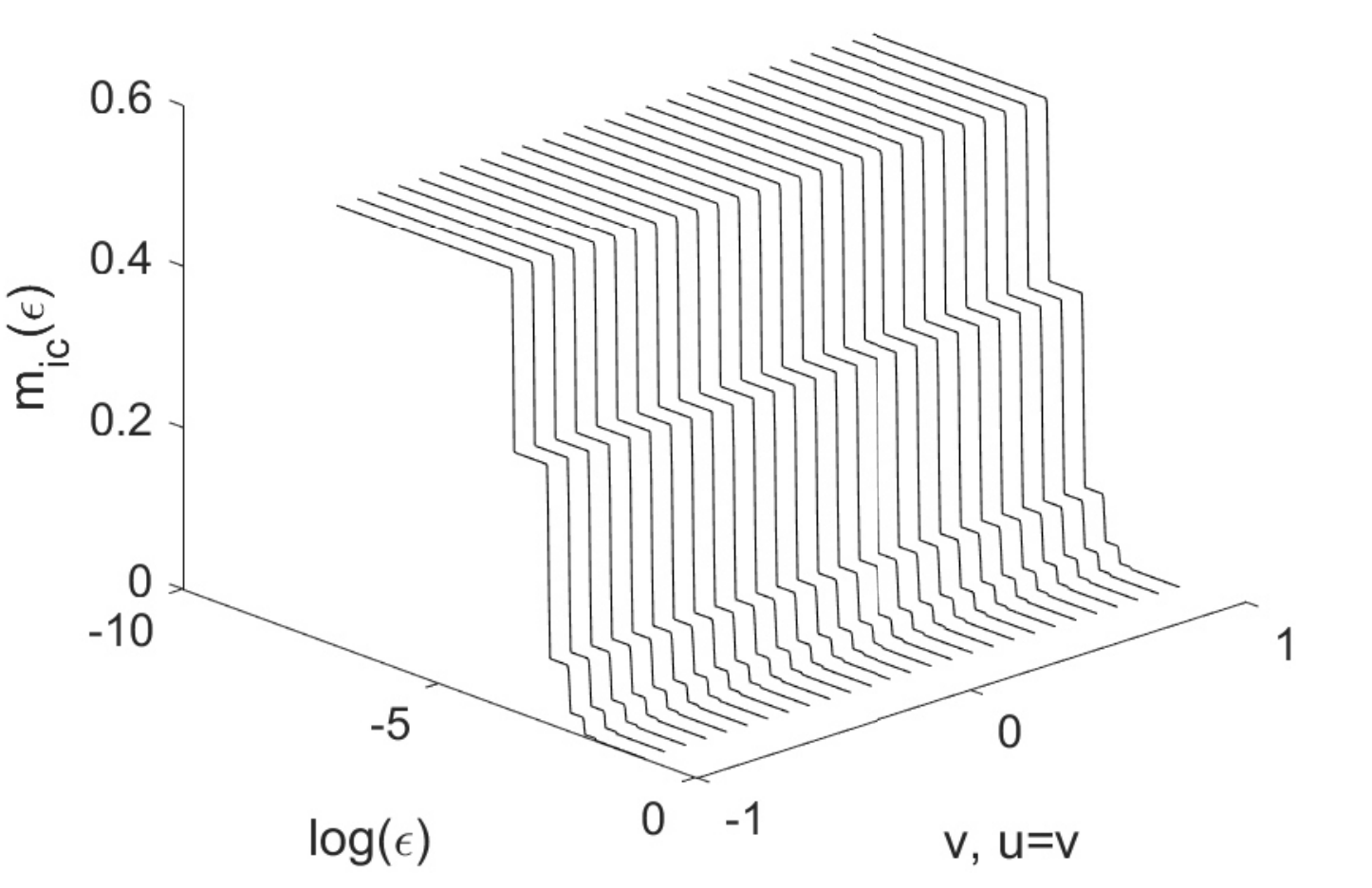} 
\includegraphics[trim = 23mm 102mm 40mm 100mm,clip, width=4.3cm, height=4cm]{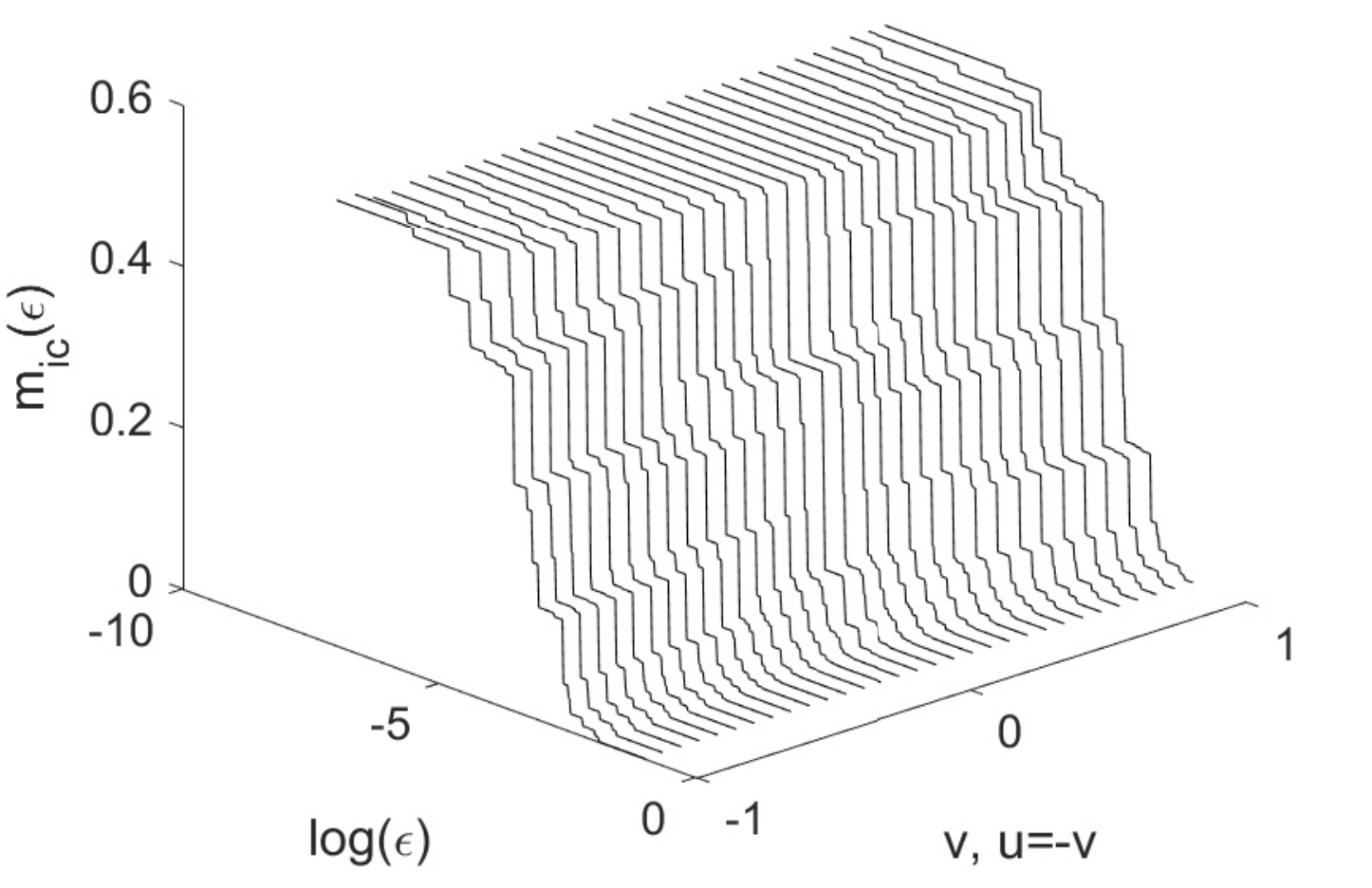} 

\hspace{1cm} (a) \hspace{4cm} (b) \hspace{4cm} (c) \hspace{4cm} (d)

\includegraphics[trim = 23mm 102mm 40mm 100mm,clip, width=4.3cm, height=4cm]{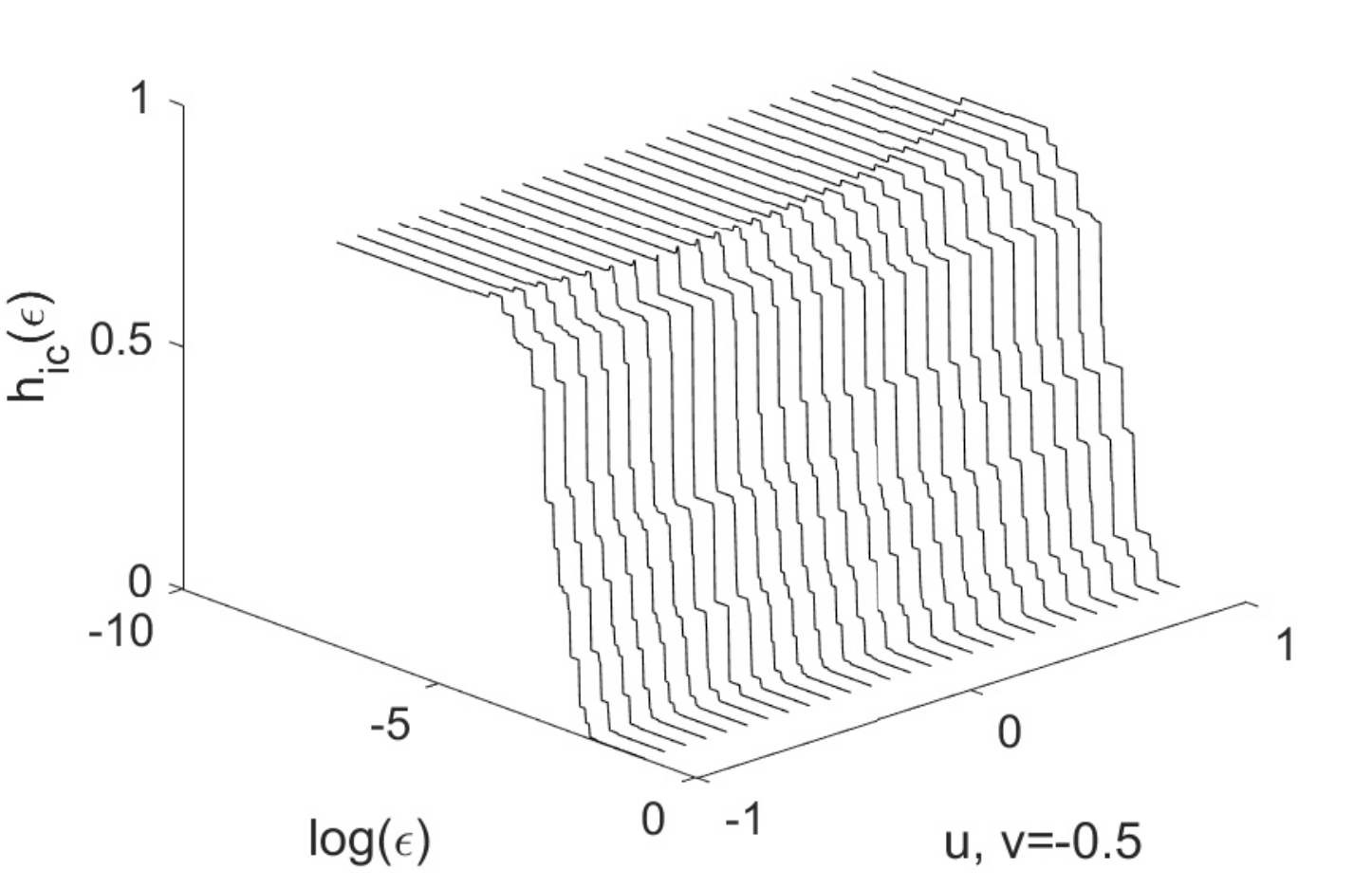} 
\includegraphics[trim = 23mm 102mm 40mm 100mm,clip, width=4.3cm, height=4cm]{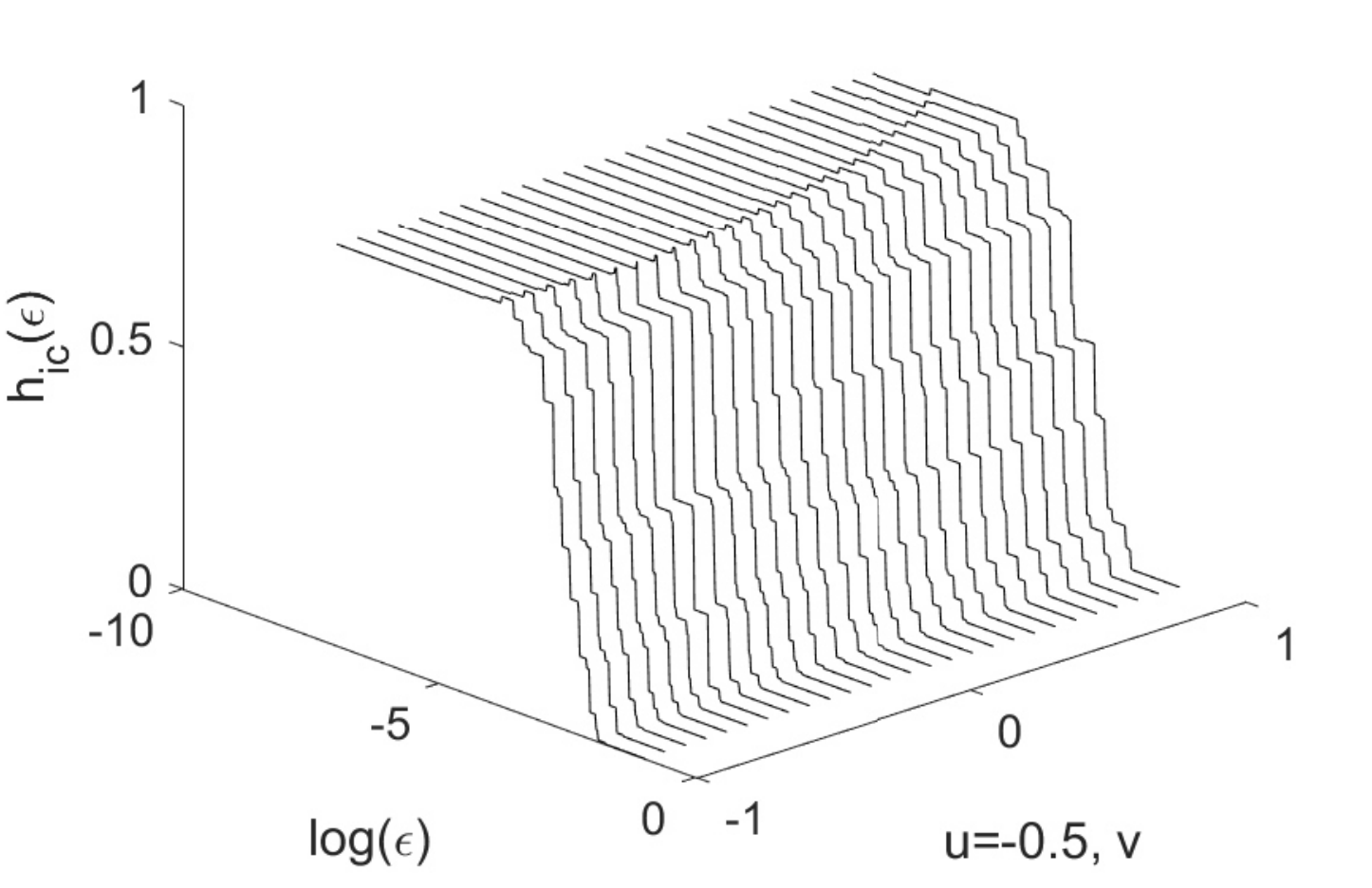} 
\includegraphics[trim = 23mm 102mm 40mm 100mm,clip, width=4.3cm, height=4cm]{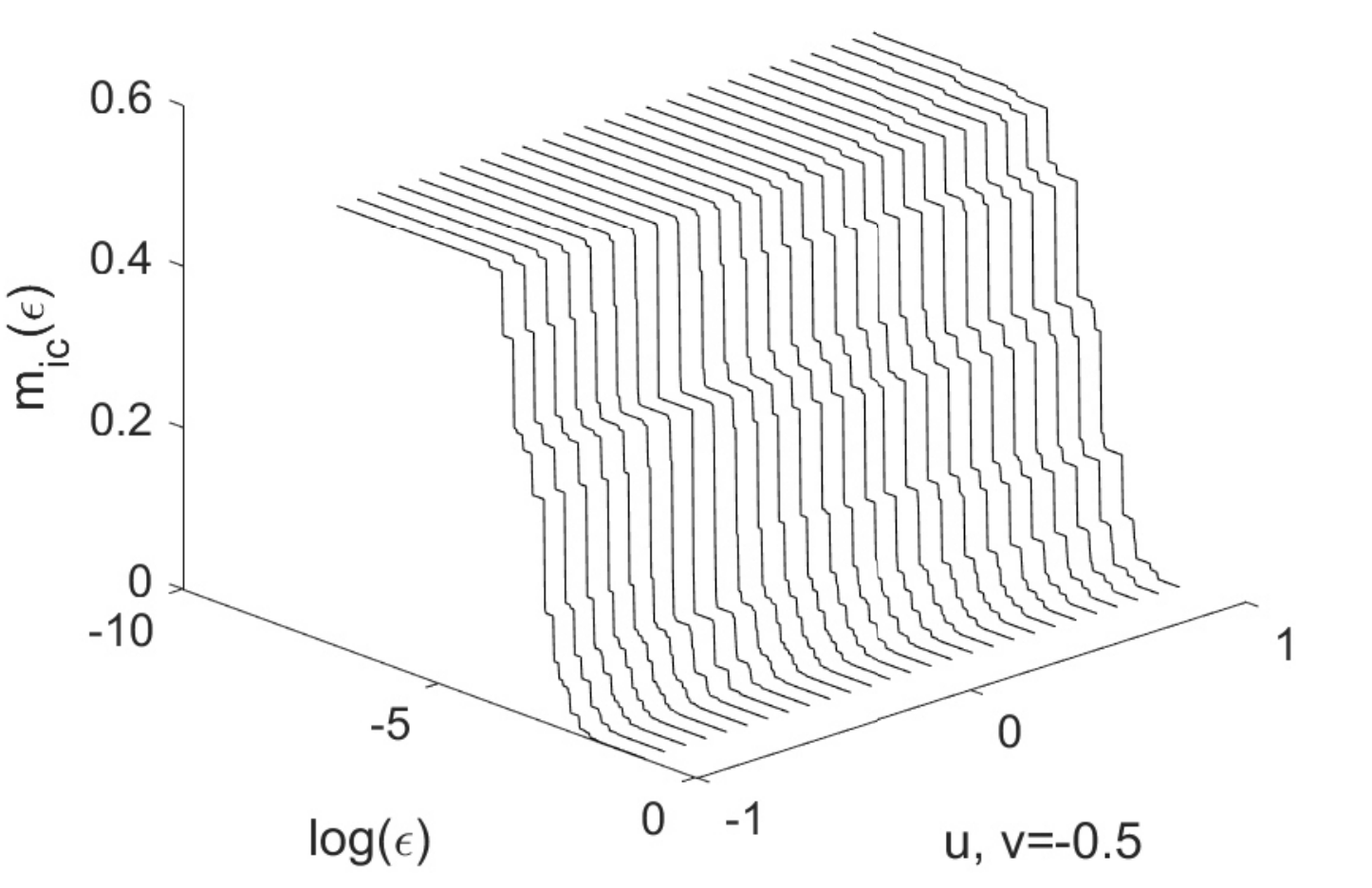} 
\includegraphics[trim = 23mm 102mm 40mm 100mm,clip, width=4.3cm, height=4cm]{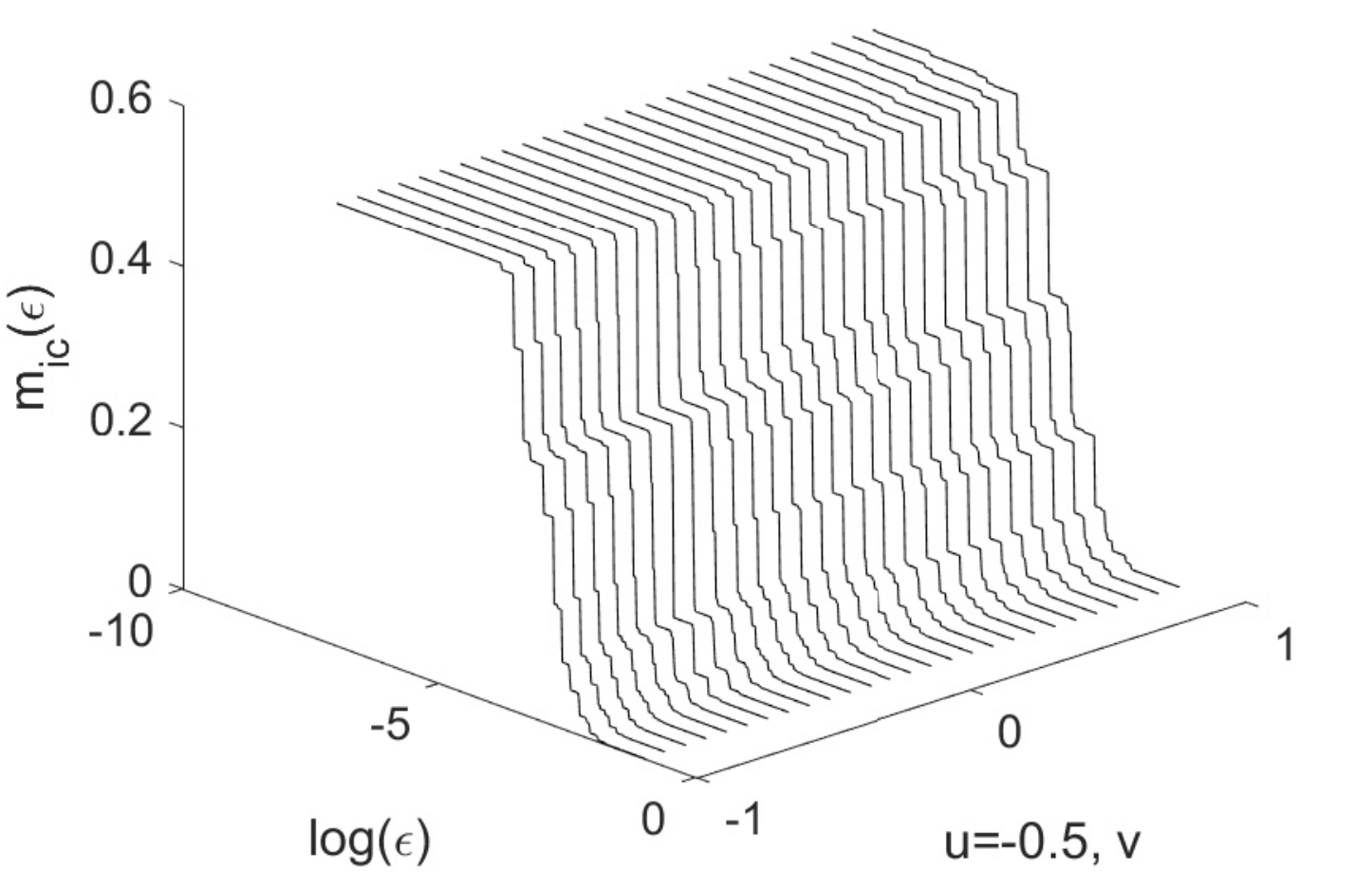} 

\hspace{1cm} (e) \hspace{4cm} (f) \hspace{4cm} (g) \hspace{4cm} (h) 

\

\caption{Information content $h_{ic}(\epsilon)$ and partial information content $m_{ic}(\epsilon)$  as a function of $\log(\epsilon)$ and  for different lines bisecting the $uv$ parameter plane. The results are for landscapes of a well--mixed population with $N=6$ and $d=5$. There are two diagonal bisections: (a),(c) $u=v$; (b),(d) $u=-v$, a vertical bisection: (e),(g) $v=-0.5$, and a horizontal bisection: (f),(h) $u=-0.5$.}
\label{fig:n1}
\end{figure}

\clearpage

\begin{figure}[t]

\includegraphics[trim = 23mm 102mm 40mm 85mm,clip, width=4.3cm, height=4cm]{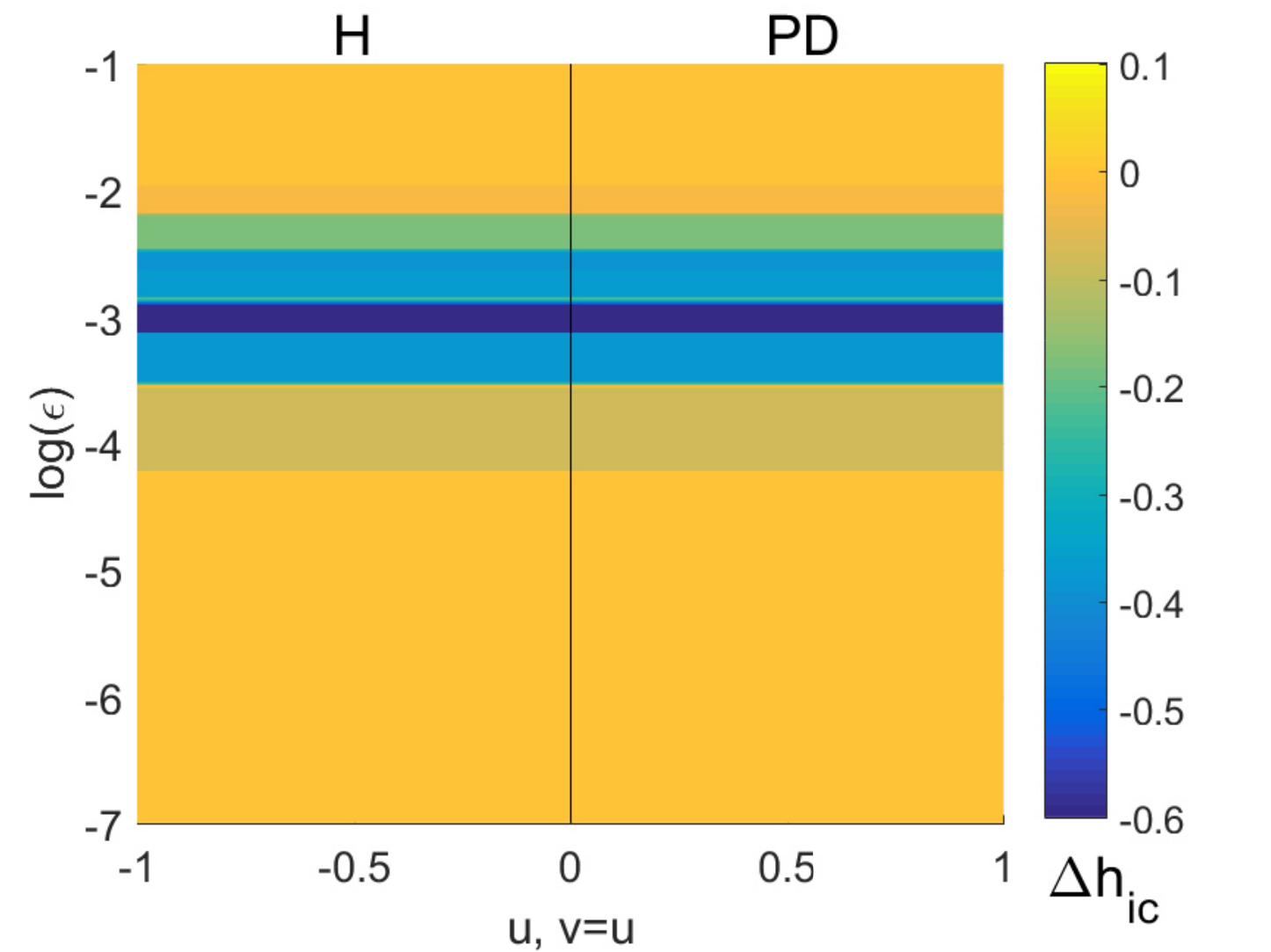} 
\includegraphics[trim = 23mm 102mm 40mm 85mm,clip, width=4.3cm, height=4cm]{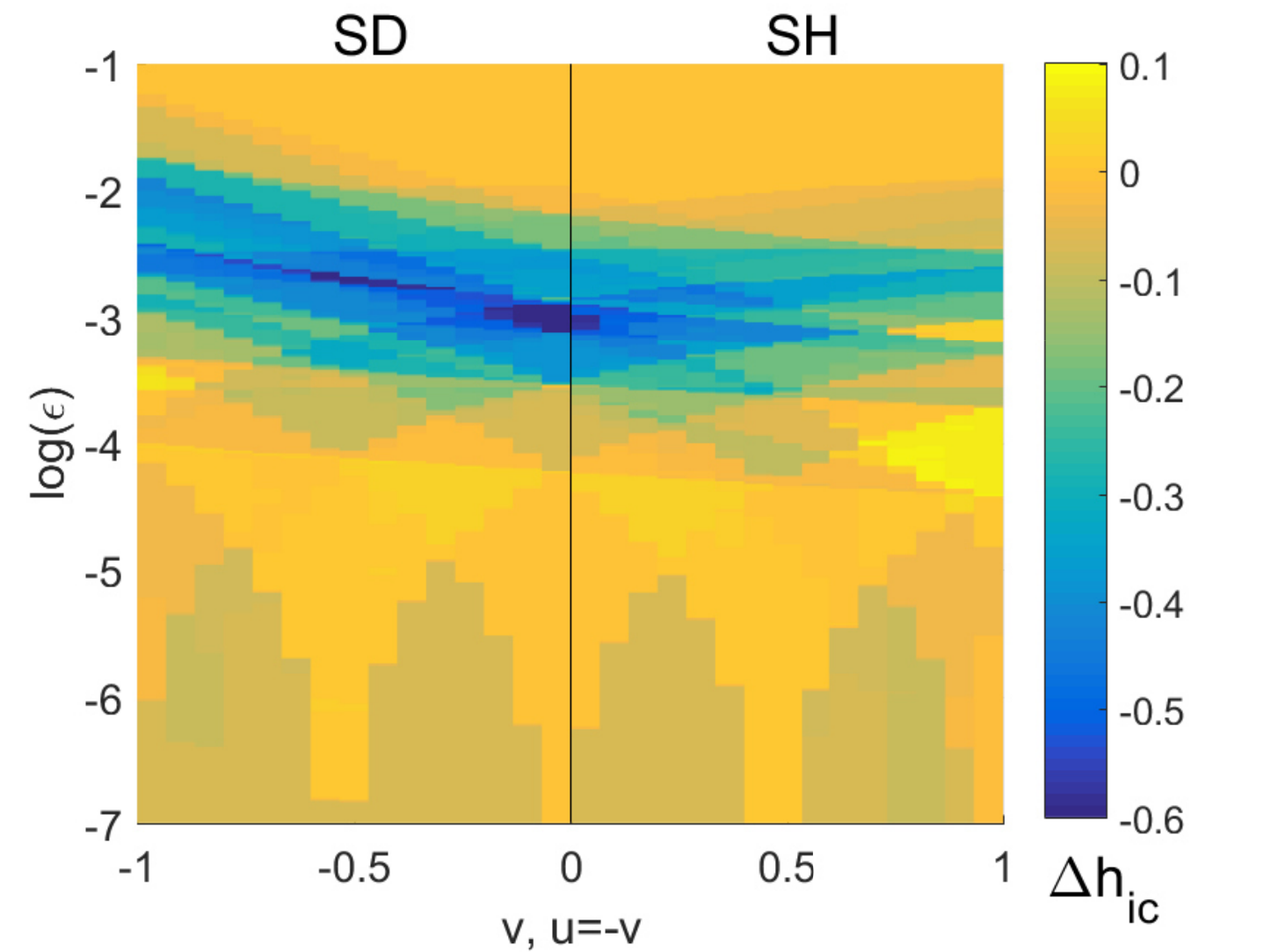} 
\includegraphics[trim = 23mm 102mm 40mm 85mm,clip, width=4.3cm, height=4cm]{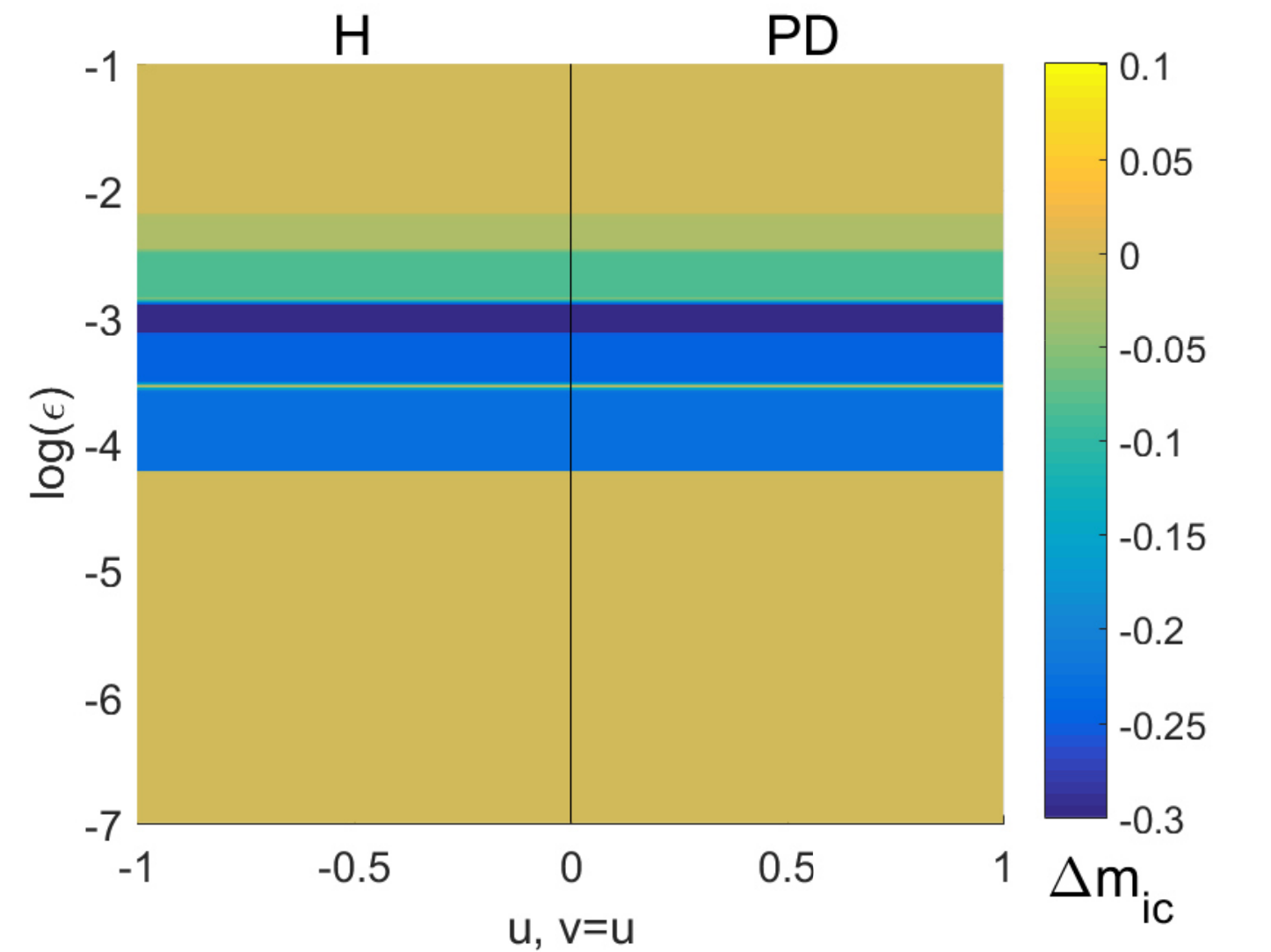} 
\includegraphics[trim = 23mm 102mm 40mm 85mm,clip, width=4.3cm, height=4cm]{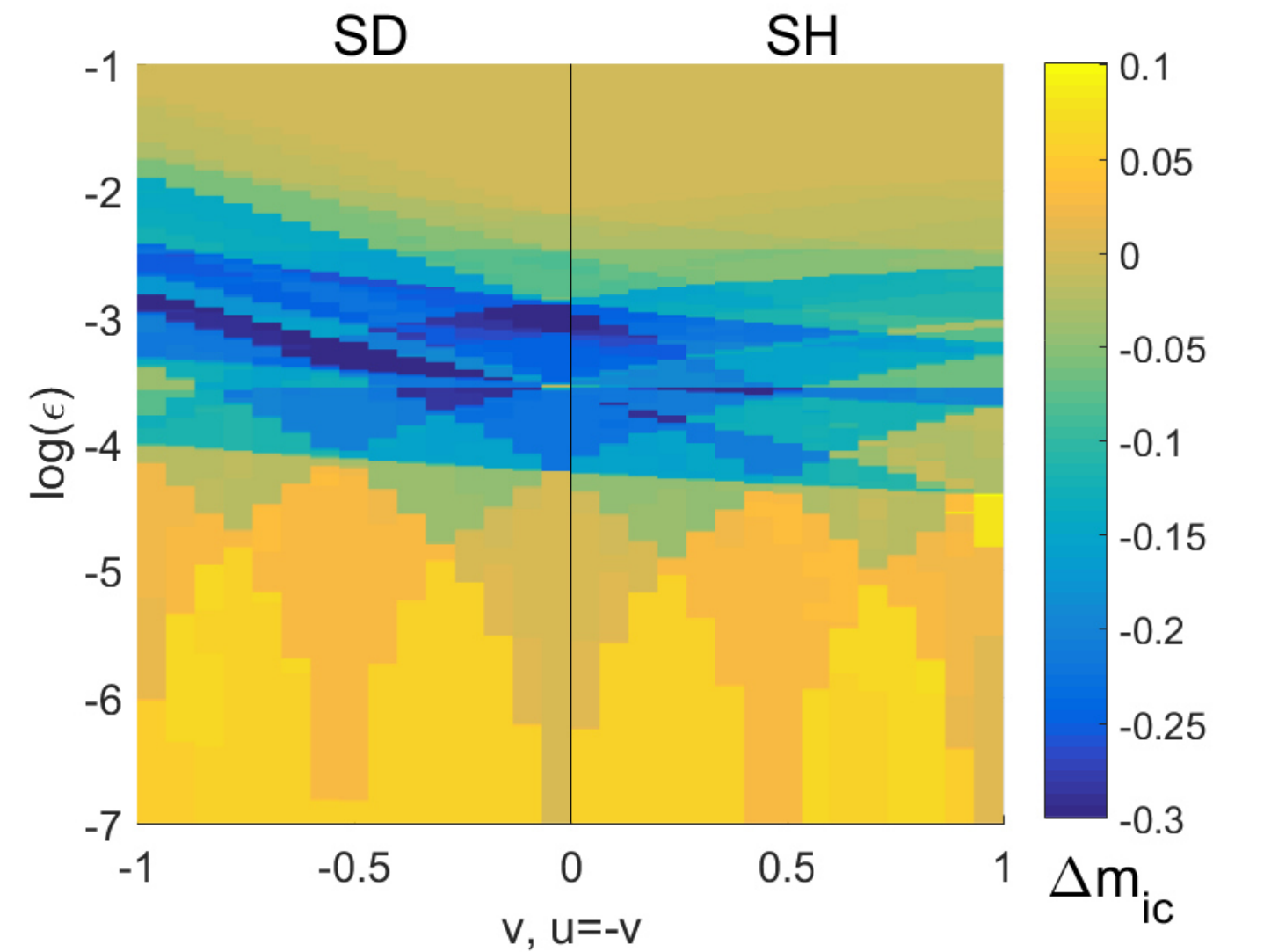} 

\hspace{1cm} (a) \hspace{4cm} (b) \hspace{4cm} (c) \hspace{4cm} (d)

\includegraphics[trim = 23mm 102mm 40mm 85mm,clip, width=4.3cm, height=4cm]{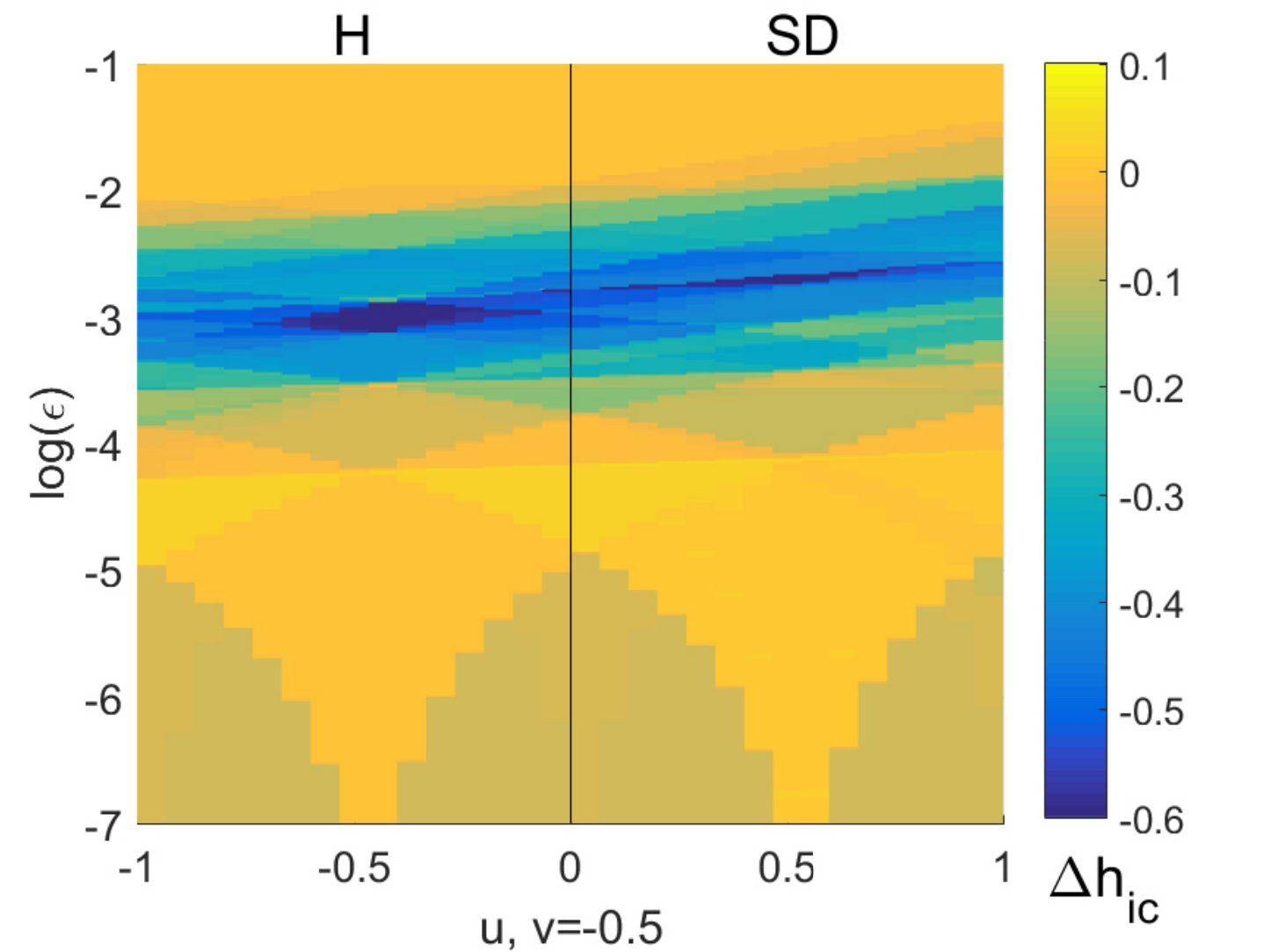} 
\includegraphics[trim = 23mm 102mm 40mm 85mm,clip, width=4.3cm, height=4cm]{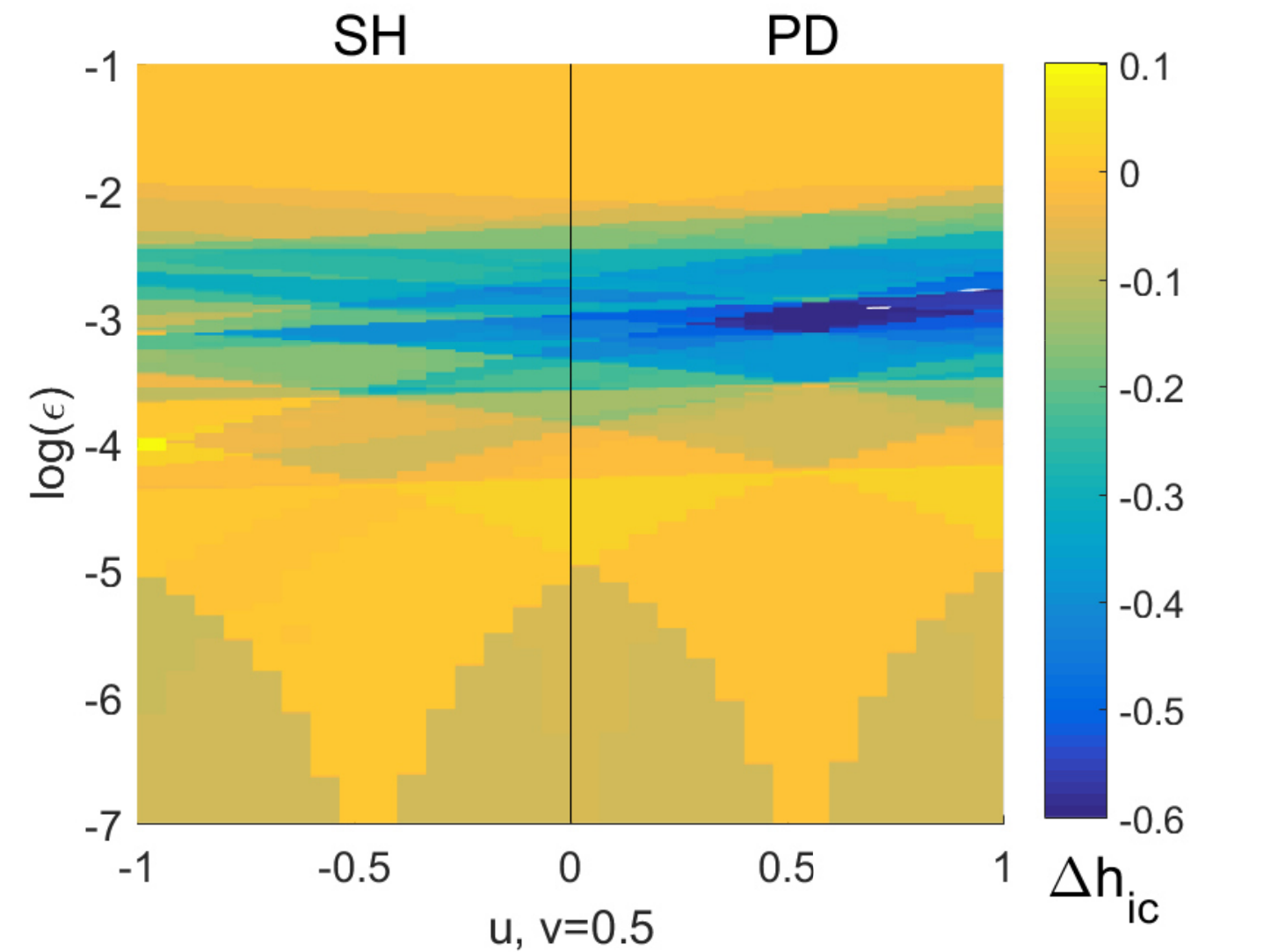} 
\includegraphics[trim = 23mm 102mm 40mm 85mm,clip, width=4.3cm, height=4cm]{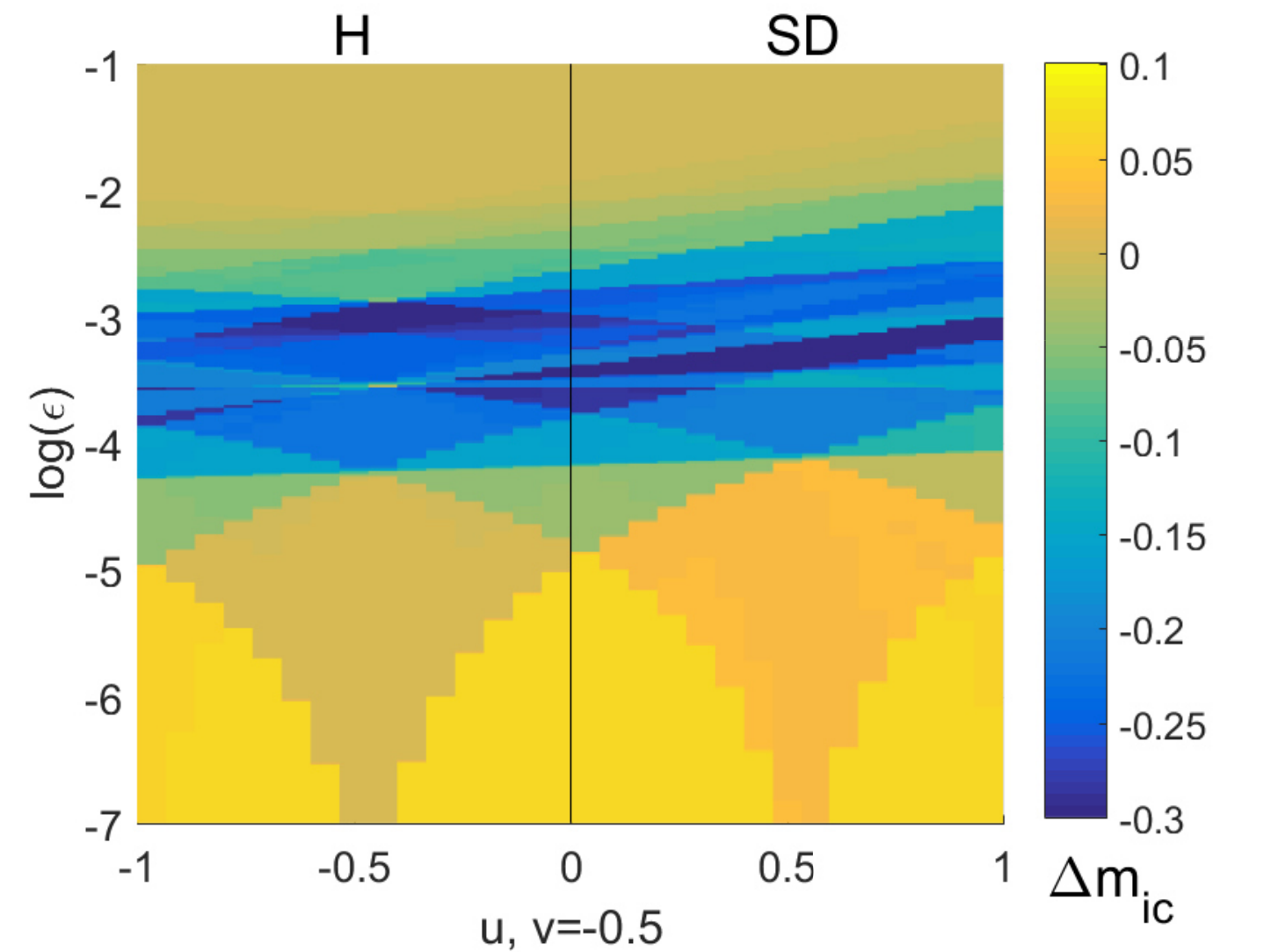} 
\includegraphics[trim = 23mm 102mm 40mm 85mm,clip, width=4.3cm, height=4cm]{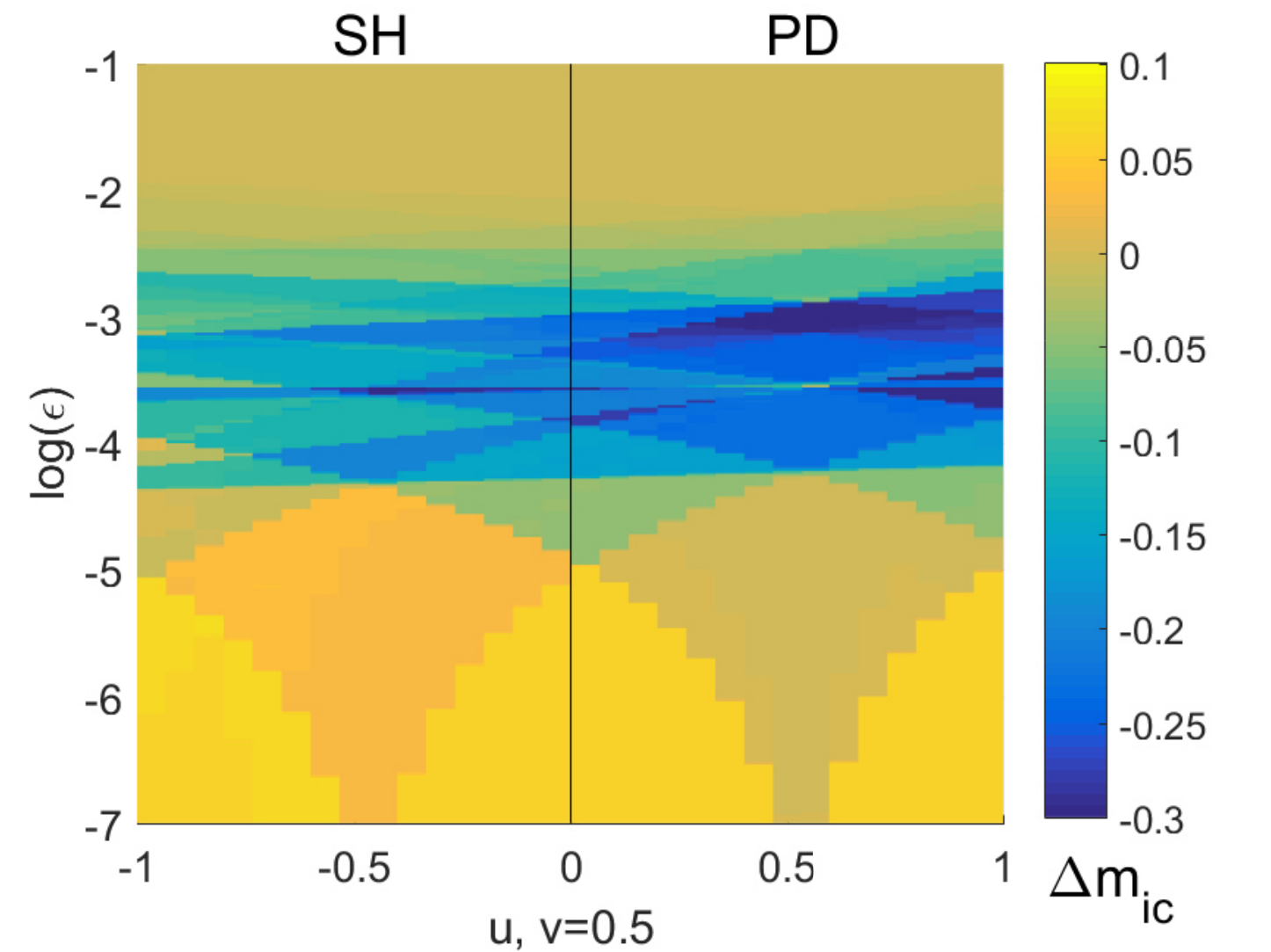} 

\hspace{1cm} (e) \hspace{4cm} (f) \hspace{4cm} (g) \hspace{4cm} (h) 

\includegraphics[trim = 23mm 102mm 40mm 85mm,clip, width=4.3cm, height=4cm]{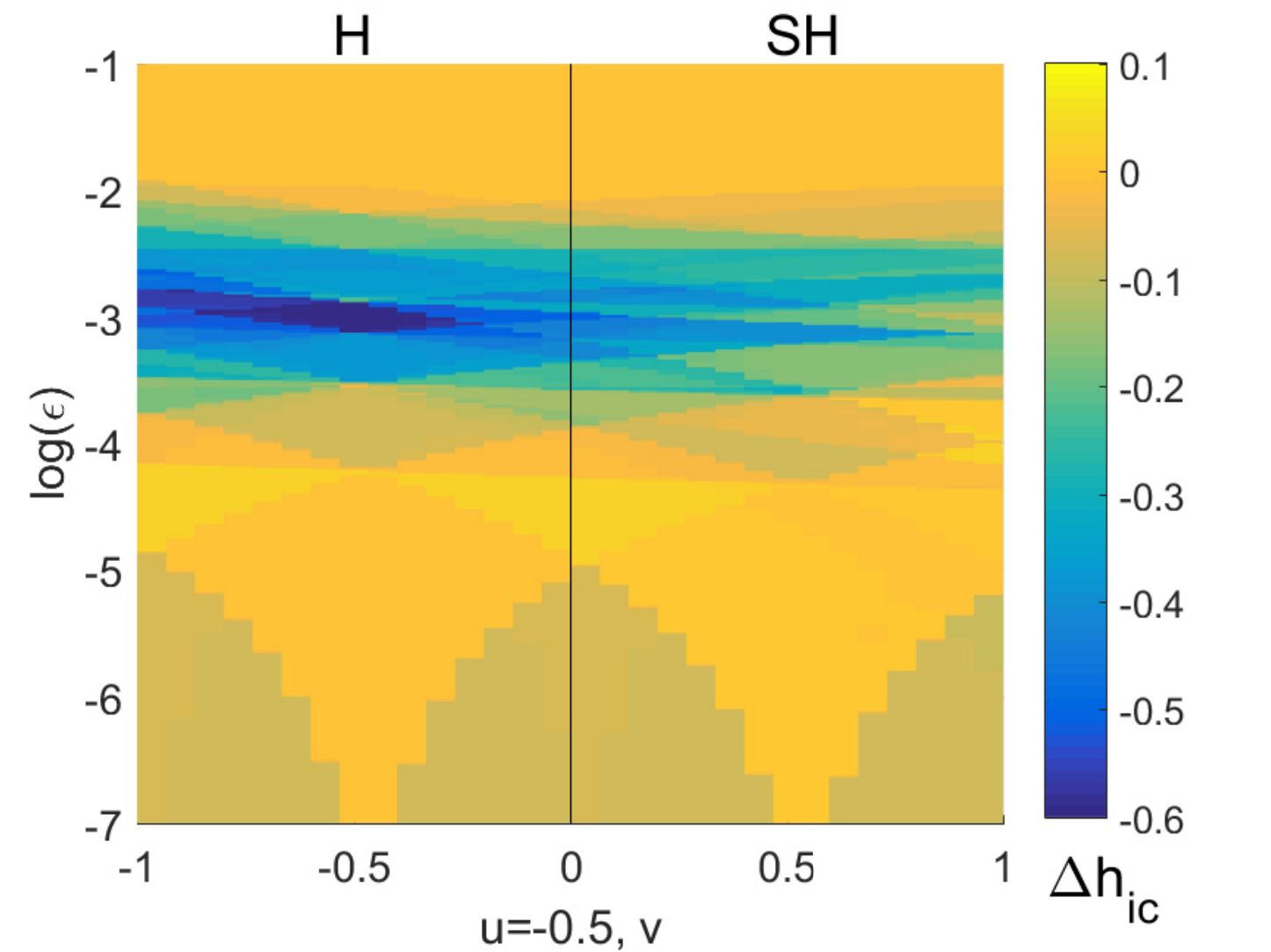} 
\includegraphics[trim = 23mm 102mm 40mm 85mm,clip, width=4.3cm, height=4cm]{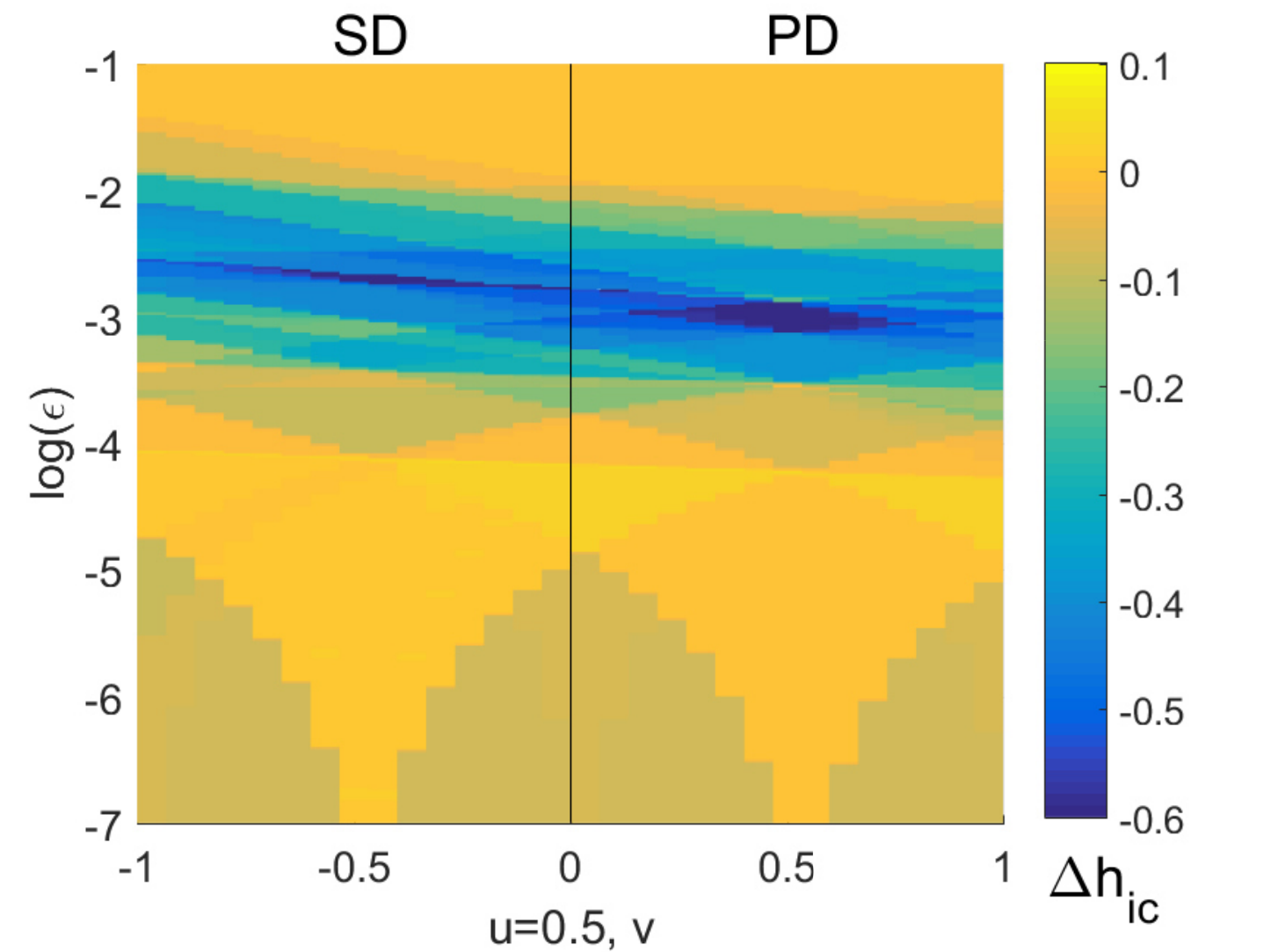} 
\includegraphics[trim = 23mm 102mm 40mm 85mm,clip, width=4.3cm, height=4cm]{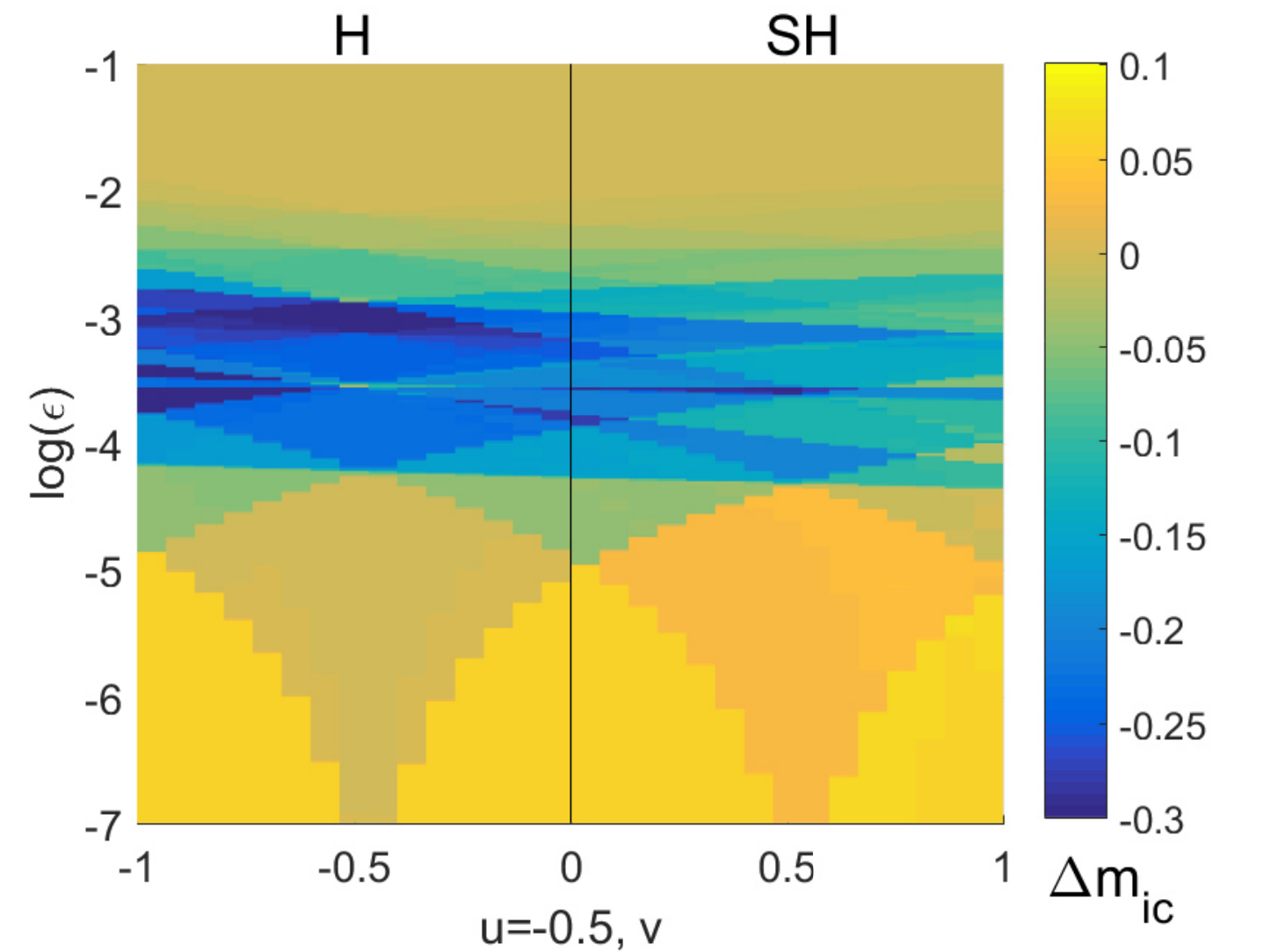} 
\includegraphics[trim = 23mm 102mm 40mm 85mm,clip, width=4.3cm, height=4cm]{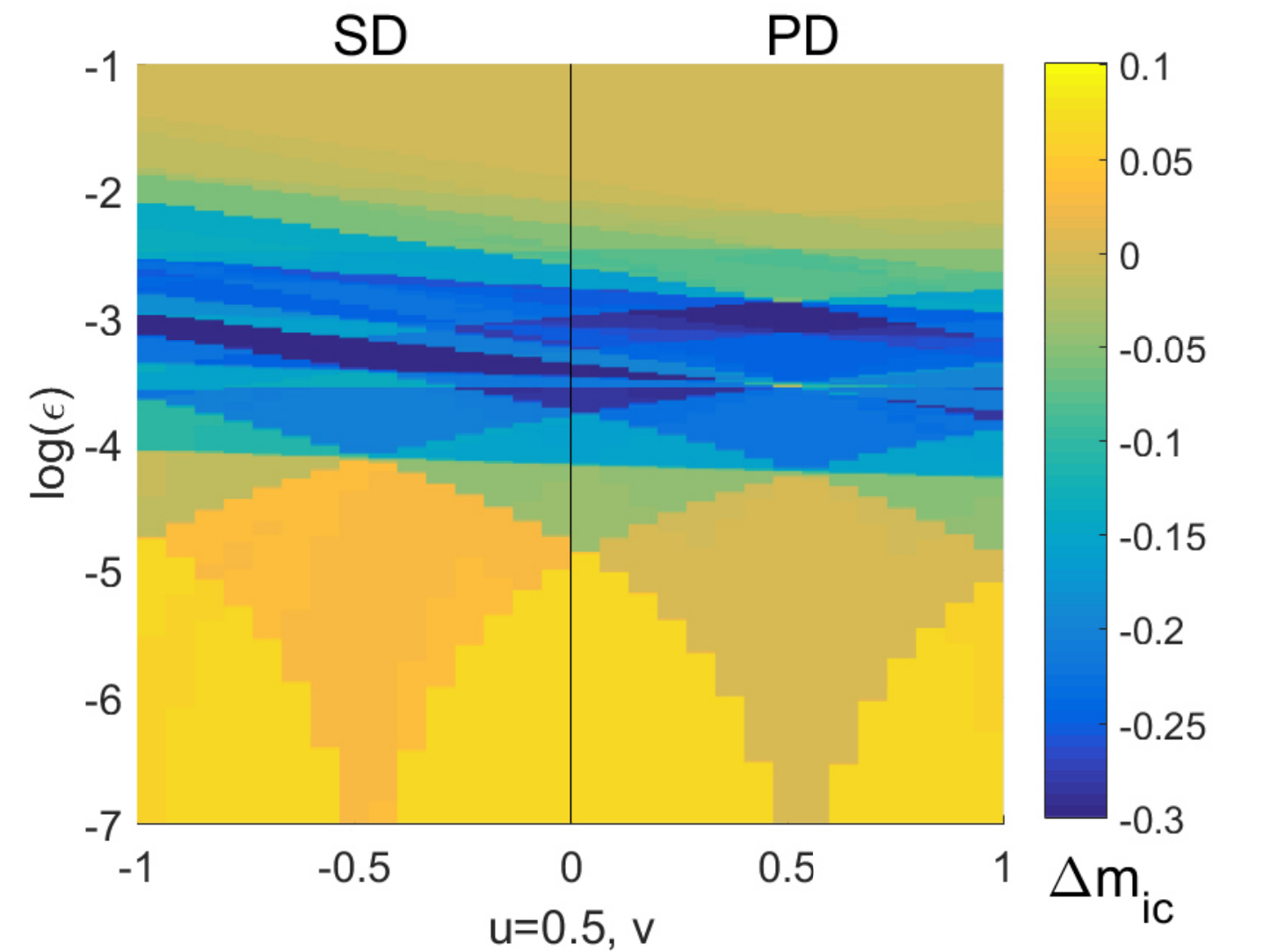} 

\hspace{1cm} (i) \hspace{4cm} (j) \hspace{4cm} (k) \hspace{4cm} (l)

\caption{Comparison between the landscapes of well--mixed and structured populations  as a function of $\epsilon$ and for different lines bisecting the $uv$ parameter plane. Above each half--plane the corresponding social dilemma is indicated, compare to Fig. \ref{fig:uv_plane}.  The difference in information content $\Delta h_{ic}=h_{ic}(\epsilon)_{str}-h_{ic}(\epsilon)_{wm}$ and partial information content $\Delta m_{ic}=m_{ic}(\epsilon)_{str}-m_{ic}(\epsilon)_{wm}$ are shown for a typical structured population with $N=6$ and $d=2$ and a well--mixed population with $N=6$ and $d=5$. There are  two diagonal bisections: (a),(c) $u=v$; (b),(d) $u=-v$, two vertical bisection: (e),(g) $v=-0.5$; (f),(h) $v=0.5$, and two horizontal bisection: (i),(k) $u=-0.5$; (j),(l) $u=0.5$.}
\label{fig:n2}
\end{figure}

\clearpage

Further results are shown in Fig. \ref{fig:n1}, which plots the information content $h_{ic}(\epsilon)$ and the partial information content $m_{ic}(\epsilon)$  against 
$\log(\epsilon)$ for different lines bisecting the $uv$ parameter plane. For $h_{ic}(\epsilon)$, we see typical curves that start from a certain level of $h_{ic}$ for $\epsilon=0$, show a small maximum for $\epsilon$ getting larger, before sharply decreasing to $h_{ic}(\epsilon)=0$. Similar curves have been reported for analyzing other landscapes~\cite{mal09,mun15,steer08,vassi00}, albeit mostly with the increase to the maximum more prominent. The results reported in Fig.  \ref{fig:n1}  have some similarity to those for  step functions~\cite{mal09} and $NK$ landscapes with a small value of epistatic interactions~\cite{vassi00}, which might be explained by these landscapes and the binary landscapes obtained for coevolutionary games mostly sharing a rather moderate ruggedness and flat areas.  This goes along with $h_{ic}(0)$ being always significantly higher than $\log_6(2)=0.3869$, which indicates that all landscapes contain some neutrality.   Also for the partial information content $m_{ic}$ we obtain  curves similar to other landscape analyses, with the largest values for $\epsilon=0$ that sink to $m_{ic}(\epsilon)=0$ for $\epsilon$ increasing. There are some further conclusions with respect to expectable game dynamics that can be drawn. For instance, 
 the landscapes across different social dilemmas are subtly different.   This becomes particularly visible by comparing the results along the different lines bisecting the $uv$--plane.  
Along the bisection $v=u$, Fig. \ref{fig:n1}a,c  traversing H and PD games, we obtain very steep descents  characterizing a rather smooth landscape, which corresponds to  game dynamics where players either all  cooperate (H) or all defect (PD). In contrast, along the bisection $v=-u$,  Fig. \ref{fig:n1}b,d we find far less steep descents, particularly close to the start and end points at $v=-1$ and $v=1$.   This means the game landscape is much more rugged, which corresponds to the more complex composition of Nash equilibria for SD and SH games. These conclusions are supported by looking at bisection that cut the $uv$--plane vertically or horizontally.  See, for instance, Fig. \ref{fig:n1}e,g  traversing H and SD games at $v=-0.5$. Again, H games ($-1<u<0$)  produce steeper descents than SD ($0<u<1$).    For the horizontal cut $u=-0.5$, Fig. \ref{fig:n1}f,h, we move from H to SH games and obtain results almost symmetric to the vertical one.  However, note the slightly different steepness at $u=1$ (Fig. \ref{fig:n1}e) as compared to $v=1$ (Fig. \ref{fig:n1}f), which may suggest that also SD and SH games  produce slight differences in the ruggedness of the underlying game  landscapes. Additional work should be done to analyze these differences. 
\begin{figure*}[t]

\includegraphics[trim = 23mm 100mm 40mm 90mm,clip, width=4.3cm, height=4cm]{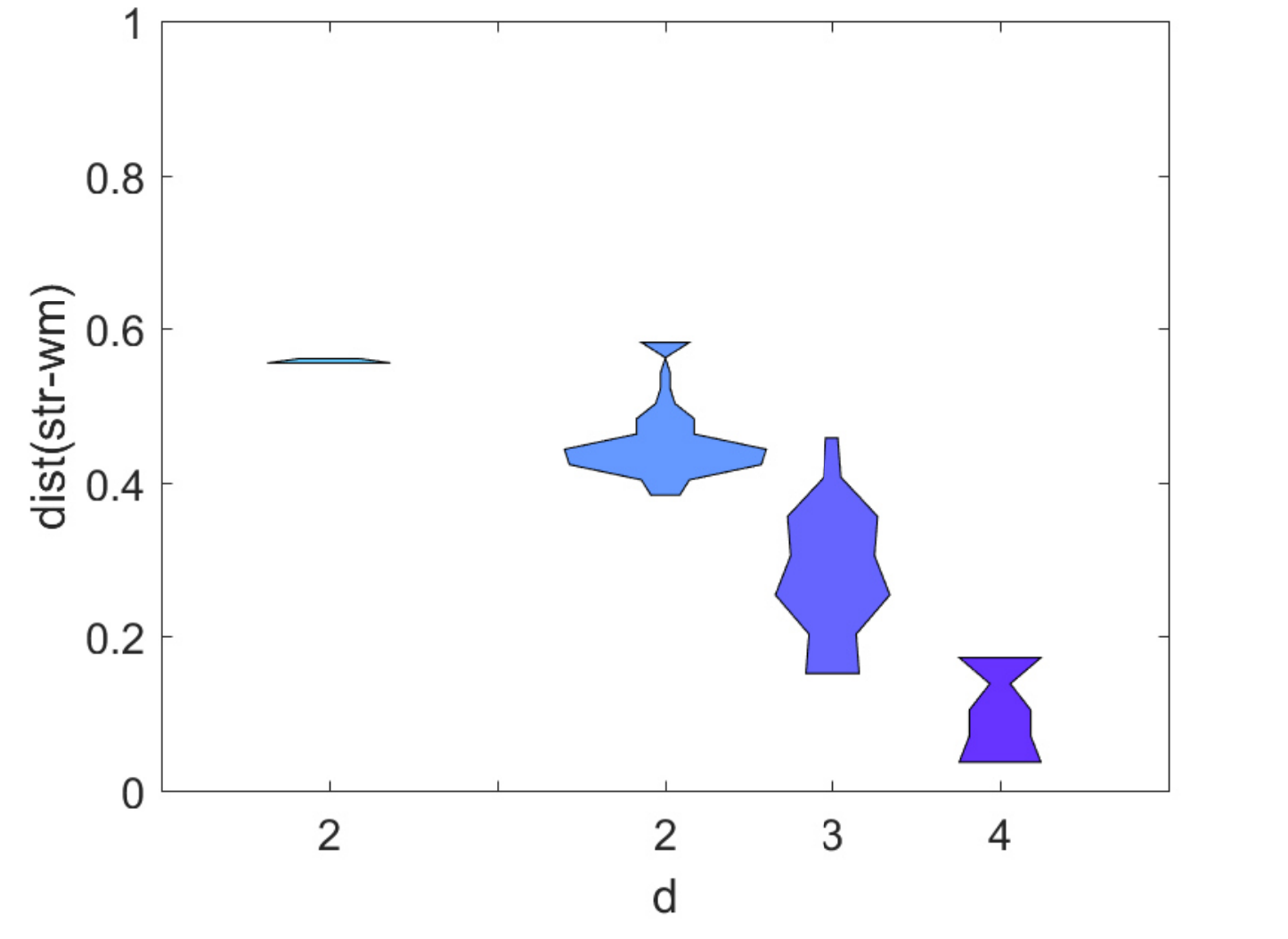} 
\includegraphics[trim = 23mm 100mm 40mm 90mm,clip, width=4.3cm, height=4cm]{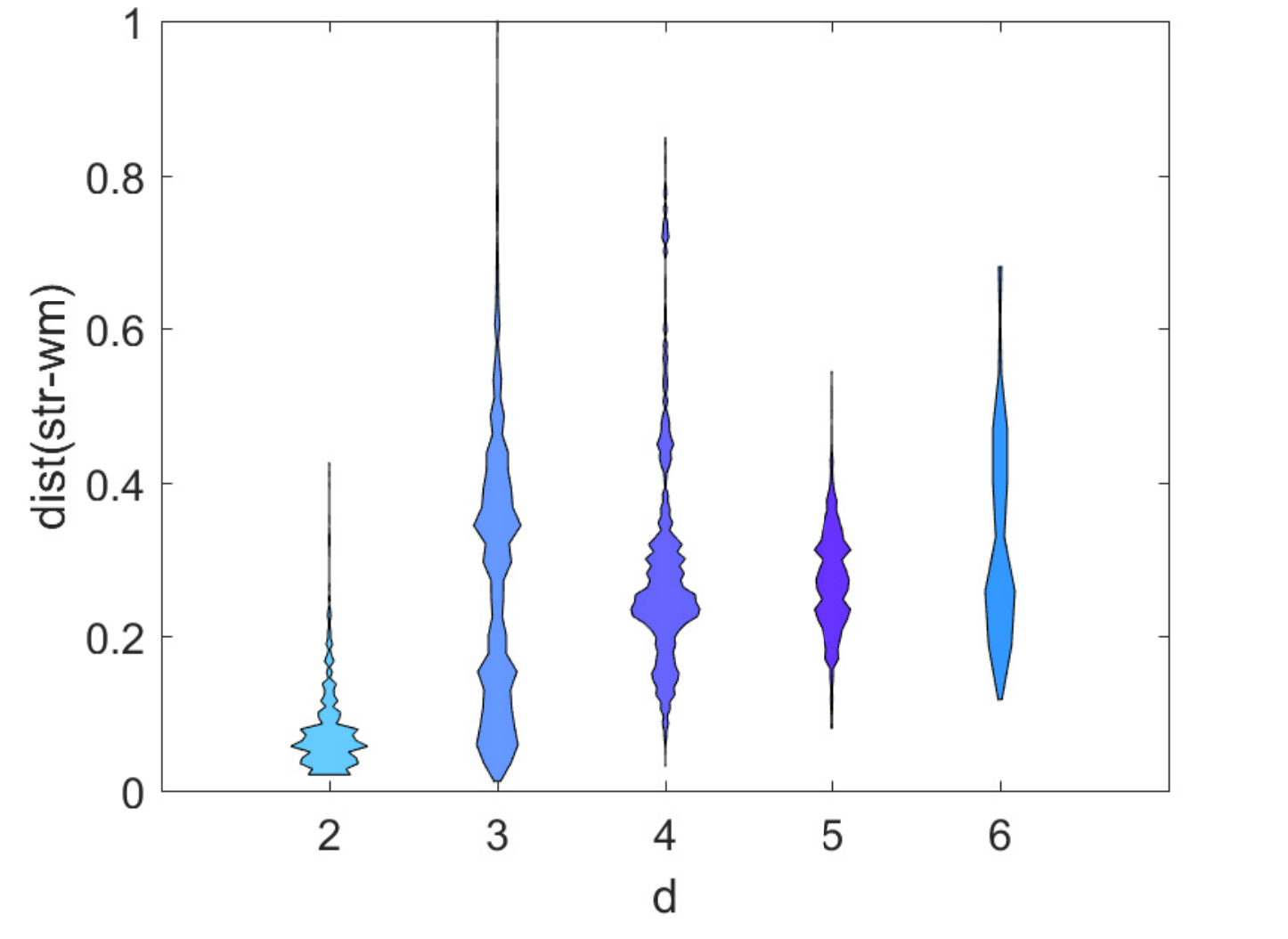} 
\includegraphics[trim = 23mm 100mm 40mm 90mm,clip, width=4.3cm, height=4cm]{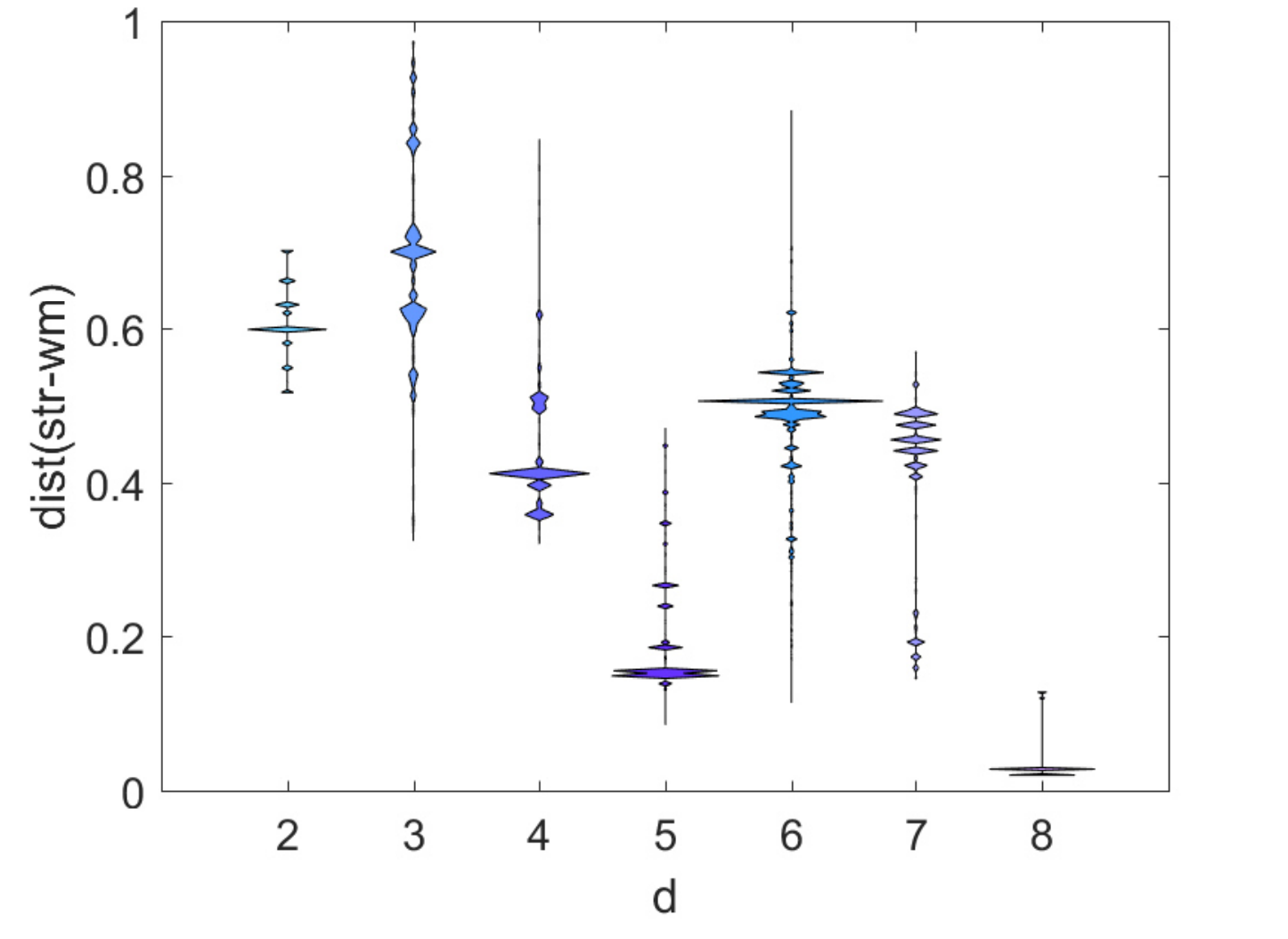} 
\includegraphics[trim = 23mm 100mm 40mm 90mm,clip, width=4.3cm, height=4cm]{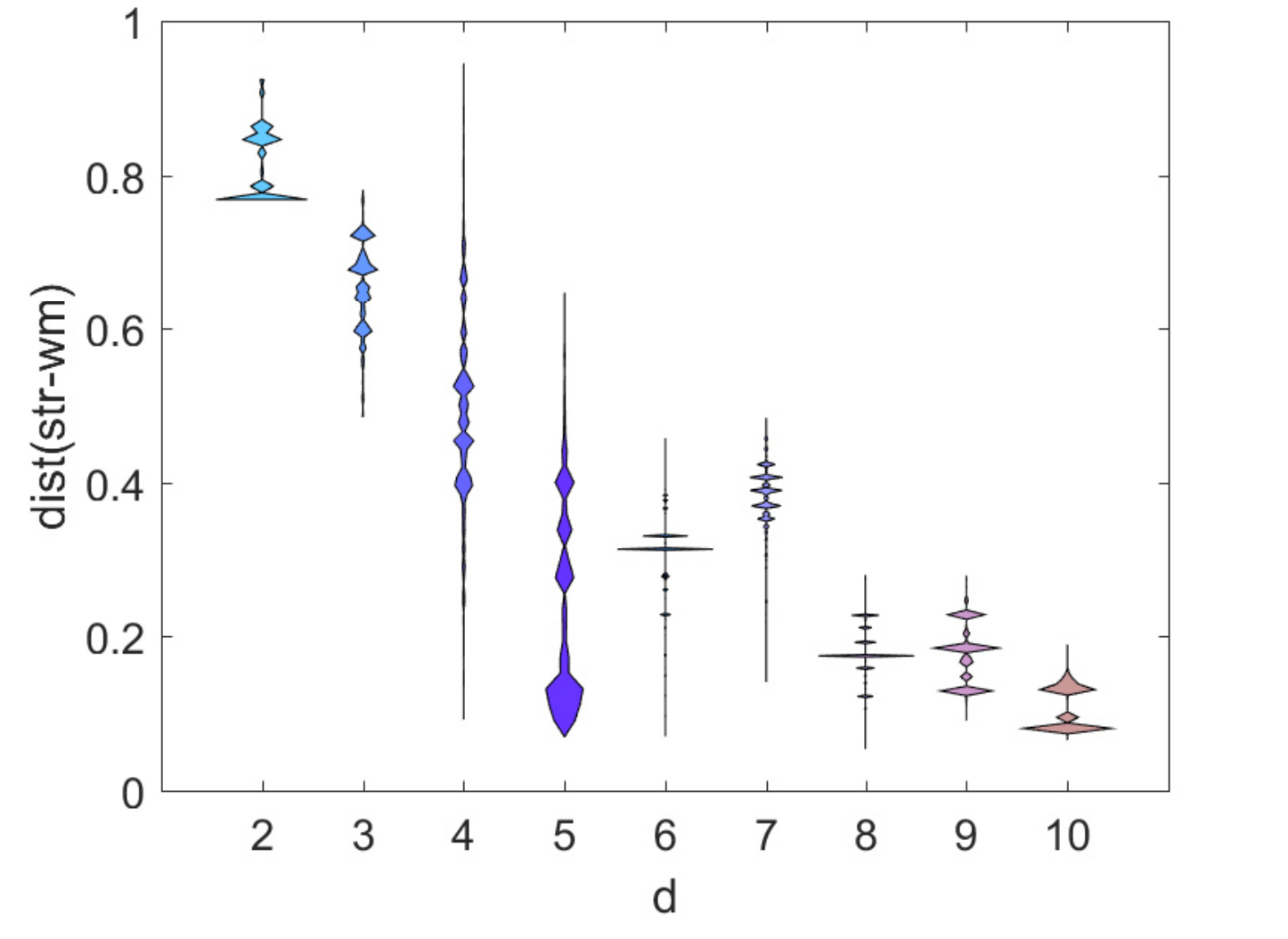} 
 
\hspace{1cm} (a) \hspace{4cm} (b) \hspace{4cm} (c) \hspace{4cm} (d) 
\caption{Violin plots of differences in information content $h_{ic}$ between well--mixed and structured populations  over interaction networks. The violins visualize the range and the distribution shape over interaction networks. The differences  are calculated over the complete square of the $uv$--plane by counting 30 equally distanced values and recording the Kullback--Leibler distance, and  for $2 \leq d\leq N-2$. (a): $N=4$ (left) and $N=6$ (right); (b): $N=8$; (c): $N=10$; (d): $N=12$. }
\label{fig:n3}
\end{figure*}

After establishing general differences and similarities between the landscape measures of structured and well--mixed populations, we next study how they vary. Fig.  \ref{fig:n2} depicts  differences in the information content and the partial information content, $\Delta h_{ic}=h_{ic}(\epsilon)_{str}-h_{ic}(\epsilon)_{wm}$ and $\Delta m_{ic}=m_{ic}(\epsilon)_{str}-m_{ic}(\epsilon)_{wm}$. The results are for   a typical  structured and a well--mixed population with $N=6$ and  either $d=2$ or $d=5$ over $\log(\epsilon)$ and different bisections of the $uv$--plane. In fact, Fig. \ref{fig:n2} views the curves in Fig. \ref{fig:n1} from atop, but now the difference between the curves for a structured and a well--mixed case are shown. Thus, the left--right half--planes in Fig. \ref{fig:n2} correspond to respective social dilemmas, as marked above each half, also compare to Fig.    \ref{fig:uv_plane}.
Generally, $h_{ic}(\epsilon)$ and $m_{ic}(\epsilon)$ are equal or slightly larger for structured as compared to well--mixed, as can be seen by the yellow, orange and ocher areas. However, for $-4.5<\log(\epsilon)<-2$, where $h_{ic}$ and $m_{ic}$ fall to zero and differences in the  descent define $\Delta h_{ic}$ and $\Delta m_{ic}$, there are green and blue areas indicating that the information content for the well--mixed population is much larger than that for the structured population. This allows to conclude that for the structured case either the fall to zero comes for smaller values of $\log(\epsilon)$ or the decent is steeper. Additional experiments (not depicted due to brevity) have shown that the differences  mainly stem from an earlier fall to zero and not from larger differences in steepness.  These results underline the substantial differences in the game landscapes over population structures, which is a major indicator for differences in game dynamics.  Another interesting result are the differences over social dilemmas, as the quantities $\Delta h_{ic}$ and $\Delta m_{ic}$ differ over $\log(\epsilon)$ and along the lines in the $uv$--plane.   An exception from this rule is the diagonal $u=v$ encompassing H and PD games, which is the same for all $-1 \leq v \leq1$ (Fig. \ref{fig:n2}a,c). This exception only applies to exactly the line $u=v$. Small perturbations to this line show again variation over $\log(\epsilon)$ and $v$. However, the differences also show a substantial degree of symmetry. This becomes particularly visible by looking at how the large negative differences in $\Delta h_{ic}$ and $\Delta m_{ic}$ (indicated by the blue and green areas) vary over $\log(\epsilon)$ and the bisections. We see that the values of $\log(\epsilon)$ fall substantially from SD to SH (Fig. \ref{fig:n2}b,d),   from  SD to H (Fig. \ref{fig:n2}e,g) and from  SD to PD (Fig. \ref{fig:n2}j,l) and more weakly from  PD to SH (Fig. \ref{fig:n2}f,h) and from  H to SH (Fig. \ref{fig:n2}i,k). In other words, they rise towards the north--west corner of the $uv$--plane and fall towards the south--east corner. 
In short, together with the south--west--to--north--east symmetry of the information content, see Fig. \ref{fig:n0}, this allows to differentiate between all 4 types of social dilemmas. Moreover, although there are  differences in how $\Delta h_{ic}$ and $\Delta m_{ic}$ vary over $\log(\epsilon)$ and bisections, they scale in a very similar manner. In other words,  $\Delta m_{ic}$ frequently gives redundant information as compared to $\Delta h_{ic}$, which may suggest that these game landscapes  are not marked by distinct flatness.  

We round off the discussion about differences in the landscapes of well--mixed and structured populations by studying the distributions over different interaction networks. Fig. \ref{fig:n3} shows violin plots over $N$ and $d$. Here, we see for $N=\{4,6,8,10,12\}$ and $2 \leq d\leq N-2$ the distribution shapes for differences between the information content $h_{ic}$. The results for $m_{ic}$ are similar and not shown due to brevity.  We measure the differences over the complete square of the $uv$--plane by counting 30 equally distanced values and recording the Kullback--Leibler distance.  
The number of bins for the histograms showing the distribution shapes in the violins is calculated according to the Freedman--Diaconis rule. We see for all $N$ and $d$ variety over different interaction networks. The differences are very small for $N=4$ and $d=2$, Fig. \ref{fig:n3}a, which is due to the fact that there are only  $\mathcal{L}_2(4)=3$ instances, with two of them producing the same result. For $N=\{6,8\}$, Fig. \ref{fig:n3}a,b, we get distribution shapes that cover a larger range, look rather different from normal distributions, and mostly show similar frequencies over the range of the histogram  with small tails at both ends, particularly for $N=8$. For $N=\{10,12\}$, Fig. \ref{fig:n3}c,d, the distribution shapes  are again different. We get frequencies that change irregularly with large outburst next to constrictions   over the range of the histogram. A possible explanation for these results is that the distributions recorded are for just a tiny and not sufficiently
representative sample of all interaction networks. While for $N=\{4,6,8\}$ all or almost all interaction networks are tested, the bound $G=8.500$ becomes small compared to  $\mathcal{L}_d(10)$ and  $\mathcal{L}_d(12)$.
 The constrictions may simply stand for interaction networks that are not in the test set.
  Further work is needed to understand how the landscapes for a well--mixed population (where all players interact with all other players) differs from one for a structured population (where there are restriction as to who--interacts-with--whom) and also how the variety changes over interaction networks, and players and coplayers.

\section{Summary and conclusions} \label{sec:con}

In this paper, an analysis of coevolutionary  game landscapes by their information content has been presented. This involves studying the landscapes across a parameter plane encompassing a continuum of different social dilemmas, including frequently studied games such as  prisoner's dilemma (PD), snowdrift (SD), stag hunt (SH), or harmony (H). The study additionally considers different interaction networks modeled by $d$--regular graphs. It was shown that the information content allows to differentiate between  different social dilemmas and also between well--mixed  and structured populations. 
For well--mixed populations there is a substantial amount of work on fixation properties and the conditions of cooperation, but 
there is also ample evidence that the evolutionary dynamics differs over the graph structure of interaction networks for structured populations, and particularity from the complete graph representing a well--mixed population~\cite{broom10,hinder15,sha12}. The landscape analysis proposed here  offers another approach to study the question of  whether and how structured and well--mixed populations vary and may therefore provide an explanatory framework for
 experimental and theoretical results on fixation and cooperation. 
 
 A next step is to link the results of the landscape analysis to fixation properties. This has been done for one typical PD and one typical SD game~\cite{rich17} and showed substantial correlations between landscape measures and fixation probability and  time. However, such experiments are somehow impeded by the fact that the computational costs of calculating fixation probabilities grow exponentially with the number of players~\cite{hinder16}. A possible remedy and alternative to a direct computation might be an approximation recently proposed~\cite{chen16} that allows calculation in polynomial time.


\end{document}